\documentclass[twocolumn,times]{aastex6}
\usepackage{amsmath}
\usepackage[nolist]{acronym}

\newcommand{\reftablefootprints}{2}
\newcommand{\reftablefollowup}{3}

\providecommand{\svnid}[1]{}

\begin{document}

\title{Supplement: Localization and broadband follow-up of the gravitational-wave transient GW150914}

\slugcomment{The Astrophysical Journal Supplement Series, 225:8, 2016 July}
\received{2016 April 27}
\accepted{2016 May 4}
\published{2016 July 20}

\defcitealias{GCN18330}{18330}
\defcitealias{GCN18331}{18331}
\defcitealias{GCN18332}{18332}
\defcitealias{GCN18333}{18333}
\defcitealias{GCN18334}{18334}
\defcitealias{GCN18335}{18335}
\defcitealias{GCN18336}{18336}
\defcitealias{GCN18337}{18337}
\defcitealias{GCN18338}{18338}
\defcitealias{GCN18339}{18339}
\defcitealias{GCN18340}{18340}
\defcitealias{GCN18341}{18341}
\defcitealias{GCN18343}{18343}
\defcitealias{GCN18344}{18344}
\defcitealias{GCN18345}{18345}
\defcitealias{GCN18346}{18346}
\defcitealias{GCN18347}{18347}
\defcitealias{GCN18348}{18348}
\defcitealias{GCN18349}{18349}
\defcitealias{GCN18350}{18350}
\defcitealias{GCN18353}{18353}
\defcitealias{GCN18354}{18354}
\defcitealias{GCN18359}{18359}
\defcitealias{GCN18361}{18361}
\defcitealias{GCN18362}{18362}
\defcitealias{GCN18363}{18363}
\defcitealias{GCN18364}{18364}
\defcitealias{GCN18370}{18370}
\defcitealias{GCN18371}{18371}
\defcitealias{GCN18372}{18372}
\defcitealias{GCN18388}{18388}
\defcitealias{GCN18390}{18390}
\defcitealias{GCN18394}{18394}
\defcitealias{GCN18395}{18395}
\defcitealias{GCN18397}{18397}
\defcitealias{GCN18420}{18420}
\defcitealias{GCN18424}{18424}
\defcitealias{GCN18474}{18474}
\defcitealias{GCN18655}{18655}
\defcitealias{GCN18690}{18690}
\defcitealias{GCN18709}{18709}
\defcitealias{GCN18851}{18851}
\defcitealias{GCN18858}{18858}
\defcitealias{GCN18903}{18903}
\defcitealias{GCN18914}{18914}
\defcitealias{GCN19013}{19013}
\defcitealias{GCN19017}{19017}
\defcitealias{GCN19021}{19021}
\defcitealias{GCN19022}{19022}
\defcitealias{GCN19034}{19034}
 \newcommand\citegcn[1]{\citetalias{GCN#1}}

\AuthorCallLimit=-1
\author{The~LIGO~Scientific~Collaboration~and~the~Virgo~Collaboration}
\author{the~Australian~Square~Kilometer~Array~Pathfinder~(ASKAP)~Collaboration}
\author{the~BOOTES~Collaboration}
\author{the~Dark~Energy~Survey~and~the~Dark~Energy~Camera~GW-EM~Collaborations}
\author{the~\emph{Fermi}~GBM~Collaboration}
\author{the~\emph{Fermi}~LAT~Collaboration}
\author{the~GRAvitational~Wave~Inaf~TeAm~(GRAWITA)}
\author{the~\emph{INTEGRAL}~Collaboration}
\author{the~Intermediate~Palomar~Transient~Factory~(iPTF)~Collaboration}
\author{the~InterPlanetary~Network}
\author{the~J-GEM~Collaboration}
\author{the~La~Silla--QUEST~Survey}
\author{the~Liverpool~Telescope~Collaboration}
\author{the~Low~Frequency~Array~(LOFAR)~Collaboration}
\author{the~MASTER~Collaboration}
\author{the~MAXI~Collaboration}
\author{the~Murchison~Wide-field~Array~(MWA)~Collaboration}
\author{the~Pan-STARRS~Collaboration}
\author{the~PESSTO~Collaboration}
\author{the~Pi~of~the~Sky~Collaboration}
\author{the~SkyMapper~Collaboration}
\author{the~\emph{Swift}~Collaboration}
\author{the~TAROT,~Zadko,~Algerian~National~Observatory,~and~C2PU~Collaboration}
\author{the~TOROS~Collaboration}
\author{the~VISTA~Collaboration}
\affil{(See the end matter for the full list of authors.)}
\email{lsc-spokesperson@ligo.org}
\email{virgo-spokesperson@ego-gw.eu}

\shorttitle{Supplement: Localization and broadband follow-up of GW150914}
\shortauthors{Abbott et al.}


\begin{abstract}
This Supplement provides supporting material for \citet{GW150914-EMFOLLOW}. We briefly summarize past \ac{EM} follow-up efforts as well as the organization and policy of the current \ac{EM} follow-up program. We compare the four probability sky maps produced for the gravitational-wave transient GW150914, and provide additional details of the \ac{EM} follow-up observations that were performed in the different bands.
\end{abstract}

\acresetall

\acused{BAYESTAR}
\acused{LIGO}
\acused{cWB}
\acused{oLIB}
\acused{MBTA}
\acused{LIB}

\section{Past and present follow-up program}

The first gravitational-wave (\acsu{GW})-triggered \ac{EM} observations were carried out during the 2009\nobreakdashes--2010 science run of the initial LIGO and Virgo detectors \citep{S6EMmethods}, featuring real\nobreakdashes-time searches for unmodeled \ac{GW} bursts and \aclp{CBC} (\acsp{CBC}\acused{CBC}; \citealt{S6EMmethods,S6lowlatencyCBC}).
\ac{GW} candidates were identified---typically within 30 minutes---and their inferred sky locations were used to plan follow-up observations with over a dozen optical and radio telescopes on the ground plus the \emph{Swift} satellite \citep{Gehrels}.
Tiles were assigned to individual facilities to target known galaxies that were consistent with the \ac{GW} localizations and that were within the 50\,Mpc nominal \ac{BNS} detectability horizon. Eight \ac{GW} candidates were followed up.
Though none of the \ac{GW} candidates were significant enough to constitute detections and the \ac{EM} candidates found were judged to be merely serendipitous sources \citep{S6Swift,S6opticalEM}, the program demonstrated the feasibility of searching in real time for \ac{GW} transients, triggering follow-up, and analyzing \ac{GW} and \ac{EM} observations jointly.

The present program of follow-up of gravitational-wave candidates involves a large number of facilities and observer teams. Instead of centrally planning the assignment of tiles to facilities, we have set up a common \ac{EM} bulletin board for facilities and observers to announce, coordinate, and visualize the footprints and wavelength coverage of their observations.
The new program builds on the \ac{GCN}\footnote{\url{http://gcn.gsfc.nasa.gov}} system that has long been established for broadband follow-up of \acp{GRB}.
We distribute times and sky positions of event candidates via machine-readable \ac{GCN} Notices, and participating facilities communicate the results of observations via short bulletins, \ac{GCN} Circulars.
A key difference is that \ac{GRB} Notices and Circulars are instantly public, whereas GW alert Notices and follow-up Circulars currently are restricted to participating groups until the event candidate in question has been published.
After four high\nobreakdashes-confidence \ac{GW} events have been published, further high-confidence \ac{GW} event candidates will be promptly released to the public.

\section{Comparison of Gravitational-Wave Sky Maps}

In the main Letter \citep{GW150914-EMFOLLOW}, we introduced four \ac{GW} sky maps produced with different methods: cWB \citep{cWB-overview}, LIB \citep{oLIB}, BAYESTAR \citep{BAYESTAR}, and LALInference \citep{LALInference}. cWB and LIB treat the \ac{GW} signal as an unmodeled burst; BAYESTAR and LALInference assume that the source is a \ac{CBC}. The LALInference sky map should be regarded as the authoritative one for this event. Table~\ref{tab:map descriptions} shows that the areas of the 10\%, 50\%, and 90\% confidence regions vary between the algorithms.
For this event, \ac{cWB} produces smaller confidence regions than the other algorithms.
While \ac{cWB} produces reasonably accurate maps for typical \ac{BBH} signals, it can systematically misestimate the sizes of large confidence regions \citep{BF2Y}.
The other algorithms are self-consistent even in this regime.
Only the LALInference results account for calibration uncertainty (systematic errors in the conversion of the photocurrent into the \ac{GW} strain signal).
Because systematic errors in the calibration phase affect the measured arrival times at the detectors, the main effect is to broaden the position uncertainty relative to the other sky maps.

Table~\ref{tab:map descriptions} also shows the intersections of the 90\% confidence regions as well as the fidelity $F(p,q) = \int \sqrt{p q}\,\mathrm{d}\Omega \in [0, 1]$ between the two maps $p$ and $q$.
All these measures show that the sky maps are similar but not identical.
Typically, this level of quantitative disagreement is distinguishable by eye and has been observed in large simulation campaigns \citep{F2Y,F2Y-recolored,BF2Y} for approximately 10\%\nobreakdashes--20\% of the simulated signals.
This even includes the bi-modality of \ac{LIB}'s $\theta_\mathrm{HL}$ distribution (see the inset of Fig.~2 of the main paper),
which is associated with a degeneracy between sky location and the handedness of the binary orbit projected on the plane of the sky.
Similar features were noted for \ac{BNS} systems as well \citep{F2Y}.

\begin{deluxetable}{crrrccccc}
    \tablecaption{\label{tab:map descriptions}Description of Sky Maps}
    \tablehead{
        \colhead{} &
        \multicolumn{3}{c}{Area\tablenotemark{a}} &
        \colhead{} & 
        \multicolumn{4}{c}{Comparison\tablenotemark{b}} \\
        \cline{2-4}
        \cline{6-9}
        \colhead{} &
        \colhead{10\%} &
        \colhead{50\%} &
        \colhead{90\%} &
        \colhead{$\theta_\mathrm{HL}$\tablenotemark{c}} &
        \colhead{cWB} &
        \colhead{LIB} &
        \colhead{BSTR} &
        \colhead{LALInf} 
    }
    \startdata
        cWB & 10 & 100 & 310 & $43^{+2}_{-2}$ & ---  & 190 & 180 & 230 \\
        LIB & 30 & 210 & 750 & $45^{+6}_{-5}$ & 0.55 & ---  & 220 & 300 \\
        BSTR & 10 & 90 & 400 & $45^{+2}_{-2}$ & 0.64 & 0.56 & ---  & 360 \\
        LALInf & 20 & 150 & 630 & $46^{+3}_{-3}$ & 0.60 & 0.57 & 0.90 & --- \\
    \enddata
    \tablenotetext{a}{Area of credible level (deg$^2$). Note that the LALInference area is consistent with but not equal to the number reported in \citet{GW150914-PARAMESTIM} due to minor differences in sampling and interpolation.}
    \tablenotetext{b}{Fidelity (below diagonal) and the intersection in deg$^2$ of the 90\% confidence regions (above diagonal).}
    \tablenotetext{c}{Mean and 10\% and 90\% percentiles of polar angle in degrees.}
\end{deluxetable}
 
\section{Gamma-ray and X-ray Observations}

The \textit{Fermi} \acl{GBM} (\acsu{GBM}; \citealt{Fermi-GBM}),  \textit{INTEGRAL} \citep{INTEGRAL}, and the \acl{IPN}~(\acsu{IPN}; \citealt{IPN}) searched for prompt high-energy emission temporally coincident with the \ac{GW} event.
Although no \ac{GRB} in coincidence with GW150914 was reported, an offline analysis of the \textit{Fermi} \ac{GBM} (8\,keV\nobreakdashes--40\,MeV) data revealed a weak transient with a duration of $\sim 1$\,s \citep{GW150914_GBM}.
A similar analysis was performed for the instruments on board \textit{INTEGRAL} \citep{INTEGRAL}, particularly the spectrometer's anticoincidence shield \citep[SPI\nobreakdashes--ACS,][75\,keV\nobreakdashes--1\,MeV]{INTEGRAL-SPI} \footnote{\textit{INTEGRAL}'s coded-mask imager \citep[IBIS,][20\nobreakdashes--200\,keV]{INTEGRAL-IBIS} was pointed far outside the \ac{GW} localization region.}.
No significant signals were detected, setting upper limits on the hard X-ray fluence at the time of the event \citep{GW150914_INTEGRAL}.
Data from the six-spacecraft, all-sky, full-time monitor \ac{IPN}, (\textit{Odyssey}--HEND, \textit{Wind}--Konus, \textit{RHESSI}, \textit{INTEGRAL}--SPI-ACS, and the \textit{Swift}--BAT\footnote{The \textit{Swift} \acl{BAT} did not intersect the GW localization at the time of the trigger}) revealed no bursts around the time of GW150914 apart from the weak \ac{GBM} signal (K. Hurley et~al. 2016, in~preparation).

The \emph{Fermi} \ac{LAT}, MAXI, and \textit{Swift} searched for high-energy afterglow emission.
The \ac{LIGO} localization first entered the \emph{Fermi} \ac{LAT} \ac{FOV} at 4200\,s after the \ac{GW} trigger and was subsequently observed in its entirety over the next 3\,hr and every 3\,hr thereafter at GeV energies \citep{GW150914_LAT}.
The entire region was also imaged in the 2\nobreakdashes--20\,keV X\nobreakdashes-ray band by the MAXI \acl{GSC} \citep{MAXI-GSC} aboard the \acl{ISS} from 86 to 77 minutes before the \ac{GW} trigger and was re-observed during each subsequent $\sim92$ minute orbit (N. Kawai et~al. 2016, in preparation).
The \textit{Swift} \acl{XRT} (\acsu{XRT}; \citealt{Swift-XRT}) followed up the \ac{GW} event starting 2.25 days after the \ac{GW} event, and covered five tiles containing eight nearby galaxies for a total $\sim$0.3\,deg$^2$ area in the 0.3--10\,keV energy range.
A 37 point tiled observation of the \acl{LMC} was executed a day later.
The \emph{Swift} \ac{UVOT} provided simultaneous ultraviolet and optical observations, giving a broadband coverage of 80\% of the \emph{Swift} \ac{XRT} \ac{FOV}.
Details of these observations are given in \citet{GW150914_Swift}.

\section{Optical and Near-IR Observations}

The optical and near-infrared observations fell into roughly two stages. During the first week, wide \ac{FOV} (1\nobreakdashes--10\,deg$^2$) telescopes tiled large areas to identify transient candidates, and then larger but narrower \ac{FOV} telescopes obtained classification spectroscopy and further photometry.
The wide \ac{FOV} instruments included DECam on the CTIO Blanco telescope \citep{DECam,DESMore}, the Kiso Wide Field Camera \citep[KWFC, J-GEM;][]{Kiso}, La Silla QUEST \citep{QUEST}, the Global MASTER Robotic Net \citep{MASTER}, the \ac{P48} as part of the \acl{IPTF} (\acsu{IPTF}; \citealt{PTF}), Pan\nobreakdashes-STARRS1 \citep{Pan-STARRS}, SkyMapper \citep{SkyMapper}, TAROT-La Silla \citep[][node of the TAROT-Zadko-Algerian National Observatory-C2PU collaboration]{TAROT}, and the \acl{VST} \citep[\acsu{VST};][GRAvitational Wave Inaf TeAm, Brocato et al. 2016, in preparation]{VST}\footnote{ESO proposal ID:095.D-0195,095.D-0079} in the optical band, and the \acl{VISTA}~(\acsu{VISTA}; \citealt{VISTA})\footnote{ESO proposal ID:095.D-0771} in the near infrared.
They represent different classes of instruments ranging in diameter from 0.25 to 4\,m and reaching apparent magnitudes from 18 to 22.5.
About one-third of these facilities followed a galaxy-targeted observational strategy, while the others tiled portions of the \ac{GW} sky maps covering  70\nobreakdashes--590\,deg$^2$.
A narrow (arcminute) \ac{FOV} facility, the 1.5\,m EABA telescope in Bosque Alegre operated by the TOROS collaboration (M.~Diaz~et~al.~2016, in preparation), also participated in the optical coverage of the \ac{GW} sky maps.
\emph{Swift} \ac{UVOT} observed simultaneously with \ac{XRT}, giving a broadband coverage of 80\% of the \emph{Swift} \ac{XRT} \ac{FOV}. 

A few tens of transient candidates identified by the wide-field telescopes were followed on the 10\,m Keck~II telescope \citep[DEIMOS;][]{Keck}, the 2\,m \acl{LT} (\acsu{LT}; \citealt{LT}), the \acl{P200} (\acsu{P200}; \citealt{P200}), the 3.6\,m ESO New Technology Telescope (within the \acl{PESSTO}, \acsu{PESSTO}; \citealt{PESSTO}), and the University of Hawaii 2.2\,m telescope (SuperNovae Integral Field Spectrograph, SNIFS). The follow-up observations of the candidate counterparts are summarized in Table~\reftablefollowup{} of the main paper.

An archival search for bright optical transients was conducted in the CASANDRA-3 all-sky camera database of BOOTES-3 \citep{BOOTES} and the all-sky survey of the Pi of the Sky telescope \citep{PiofTheSky}, both covering the entire southern sky map. The BOOTES-3 images are the only observations simultaneous to GW150914 available to search for prompt/early optical emission. They reached a limiting magnitude of 5 due to poor weather conditions (GCN~\citegcn{19022}). The Pi of the Sky telescope images were taken 12 days after GW150914 and searched for transients brighter than $R < 11.5$\,mag (GCN~\citegcn{19034}).

\section{Radio Observations}

The radio telescopes involved in the \ac{EM} follow-up program have the capability to observe a wide range of frequencies with different levels of sensitivity, and a range of \acp{FOV} covering both the northern and southern skies (Tables \reftablefootprints{} and \reftablefollowup{} of the main paper).
The \acl{LOFAR}~(\acsu{LOFAR}; \citealt{LOFAR}) and the \acl{MWA}~(\acsu{MWA}; \citealt{MWA}) are phased array dipole antennas sensitive to meter wavelengths with large \acp{FOV} ($\approx$50\,deg$^2$ with uniform sensitivity for the LOFAR observations carried out as part of this follow-up program; and up to 1200\,deg$^2$ for \ac{MWA}).
The \acl{ASKAP} (\acsu{ASKAP}; \citealt{ASKAPb}) is an interferometric array composed of 36 12\,m diameter dish antennas. The \acl{VLA} (\acsu{VLA}; \citealt{VLA}) is a 27 antenna array, with dishes of 25\,m diameter. Both \ac{ASKAP} and \ac{VLA} are sensitive from centimeter to decimeter wavelengths.

\ac{MWA} started observing $3$ days after the \ac{GW} trigger with a 30\,MHz bandwidth around a central frequency of 118\,MHz and reached an rms noise level of about 40\,mJy\,beam$^{-1}$ in a synthesized beam of about $3\arcmin$. The  \ac{ASKAP} observations used the five-element Boolardy Engineering Test Array (BETA; \citealt{ASKAPa}), which has an \ac{FOV} of $\approx$25\,deg$^{2}$ and FWHM synthesized beam of $1\arcmin-3\arcmin$. These observations were performed with a 300\,MHz bandwidth around a central frequency of 863.5\,MHz, from $\approx$7 to $\approx$14 days after the \ac{GW} trigger, reaching rms sensitivities of $1-3$\,mJy\,beam$^{-1}$.  \ac{LOFAR} conducted three observations from $\approx$7 days to $\approx$3 months following the GW trigger, reaching a rms sensitivity of $\approx$2.5\,mJy\,beam$^{-1}$ at 145\,MHz, with a bandwidth of 11.9\,MHz and a spatial resolution of $\approx$50\arcsec.
\ac{ASKAP}, \ac{LOFAR}, and \ac{MWA} all performed tiled observations aimed at covering a large area of the \ac{GW} region.

The \ac{VLA} performed follow-up observations of GW150914 from $\approx$1 to $\approx$4 months after the \ac{GW} trigger\footnote{VLA/15A-339, PI: A. Corsi}, and targeted selected candidate optical counterparts detected by \ac{IPTF}.
\ac{VLA} observations were carried out in the most compact array configuration (D~configuration) at a central frequency of $\approx$6\,GHz (primary beam FWHP of $\approx$9\arcmin, and synthesized beam FWHP of $\approx$12\arcsec).
The rms sensitivity of these \ac{VLA} observations was $\approx$8\nobreakdashes--10\,$\mu$Jy\,beam$^{-1}$.

\providecommand{\acrolowercase}[1]{\lowercase{#1}}

\begin{acronym}
\acro{2D}[2D]{two\nobreakdashes-dimensional}
\acro{2+1D}[2+1D]{2+1\nobreakdashes--dimensional}
\acro{3D}[3D]{three\nobreakdashes-dimensional}
\acro{2MASS}[2MASS]{Two Micron All Sky Survey}
\acro{AdVirgo}[AdVirgo]{Advanced Virgo}
\acro{AMI}[AMI]{Arcminute Microkelvin Imager}
\acro{AGN}[AGN]{active galactic nucleus}
\acroplural{AGN}[AGN\acrolowercase{s}]{active galactic nuclei}
\acro{aLIGO}[aLIGO]{Advanced \acs{LIGO}}
\acro{ASKAP}[ASKAP]{Australian \acl{SKA} Pathfinder}
\acro{ATCA}[ATCA]{Australia Telescope Compact Array}
\acro{ATLAS}[ATLAS]{Asteroid Terrestrial-impact Last Alert System}
\acro{BAT}[BAT]{Burst Alert Telescope\acroextra{ (instrument on \emph{Swift})}}
\acro{BATSE}[BATSE]{Burst and Transient Source Experiment\acroextra{ (instrument on \acs{CGRO})}}
\acro{BAYESTAR}[BAYESTAR]{BAYESian TriAngulation and Rapid localization}
\acro{BBH}[BBH]{binary black hole}
\acro{BHBH}[BHBH]{\acl{BH}\nobreakdashes--\acl{BH}}
\acro{BH}[BH]{black hole}
\acro{BNS}[BNS]{binary neutron star}
\acro{CARMA}[CARMA]{Combined Array for Research in Millimeter\nobreakdashes-wave Astronomy}
\acro{CASA}[CASA]{Common Astronomy Software Applications}
\acro{CFH12k}[CFH12k]{Canada--France--Hawaii $12\,288 \times 8\,192$ pixel CCD mosaic\acroextra{ (instrument formerly on the Canada--France--Hawaii Telescope, now on the \ac{P48})}}
\acro{CRTS}[CRTS]{Catalina Real-time Transient Survey}
\acro{CTIO}[CTIO]{Cerro Tololo Inter-American Observatory}
\acro{CBC}[CBC]{compact binary coalescence}
\acro{CCD}[CCD]{charge coupled device}
\acro{CDF}[CDF]{cumulative distribution function}
\acro{CGRO}[CGRO]{Compton Gamma Ray Observatory}
\acro{CMB}[CMB]{cosmic microwave background}
\acro{CRLB}[CRLB]{Cram\'{e}r\nobreakdashes--Rao lower bound}
\acro{cWB}[\acrolowercase{c}WB]{Coherent WaveBurst}
\acro{DASWG}[DASWG]{Data Analysis Software Working Group}
\acro{DBSP}[DBSP]{Double Spectrograph\acroextra{ (instrument on \acs{P200})}}
\acro{DCT}[DCT]{Discovery Channel Telescope}
\acro{DECAM}[DECam]{Dark Energy Camera\acroextra{ (instrument on the Blanco 4\nobreakdashes-m telescope at \acs{CTIO})}}
\acro{DES}[DES]{Dark Energy Survey}
\acro{DFT}[DFT]{discrete Fourier transform}
\acro{EM}[EM]{electromagnetic}
\acro{ER8}[ER8]{eighth engineering run}
\acro{FD}[FD]{frequency domain}
\acro{FAR}[FAR]{false alarm rate}
\acro{FFT}[FFT]{fast Fourier transform}
\acro{FIR}[FIR]{finite impulse response}
\acro{FITS}[FITS]{Flexible Image Transport System}
\acro{FLOPS}[FLOPS]{floating point operations per second}
\acro{FOV}[FOV]{field of view}
\acroplural{FOV}[FOV\acrolowercase{s}]{fields of view}
\acro{FTN}[FTN]{Faulkes Telescope North}
\acro{FWHM}[FWHM]{full width at half-maximum}
\acro{GBM}[GBM]{Gamma-ray Burst Monitor\acroextra{ (instrument on \emph{Fermi})}}
\acro{GCN}[GCN]{Gamma-ray Coordinates Network}
\acro{GMOS}[GMOS]{Gemini Multi-object Spectrograph\acroextra{ (instrument on the Gemini telescopes)}}
\acro{GRB}[GRB]{gamma-ray burst}
\acro{GSC}[GSC]{Gas Slit Camera}
\acro{GSL}[GSL]{GNU Scientific Library}
\acro{GTC}[GTC]{Gran Telescopio Canarias}
\acro{GW}[GW]{gravitational wave}
\acro{HAWC}[HAWC]{High\nobreakdashes-Altitude Water \v{C}erenkov Gamma\nobreakdashes-Ray Observatory}
\acro{HCT}[HCT]{Himalayan Chandra Telescope}
\acro{HEALPix}[HEALP\acrolowercase{ix}]{Hierarchical Equal Area isoLatitude Pixelization}
\acro{HEASARC}[HEASARC]{High Energy Astrophysics Science Archive Research Center}
\acro{HETE}[HETE]{High Energy Transient Explorer}
\acro{HFOSC}[HFOSC]{Himalaya Faint Object Spectrograph and Camera\acroextra{ (instrument on \acs{HCT})}}
\acro{HMXB}[HMXB]{high\nobreakdashes-mass X\nobreakdashes-ray binary}
\acroplural{HMXB}[HMXB\acrolowercase{s}]{high\nobreakdashes-mass X\nobreakdashes-ray binaries}
\acro{HSC}[HSC]{Hyper Suprime\nobreakdashes-Cam\acroextra{ (instrument on the 8.2\nobreakdashes-m Subaru telescope)}}
\acro{IACT}[IACT]{imaging atmospheric \v{C}erenkov telescope}
\acro{IIR}[IIR]{infinite impulse response}
\acro{IMACS}[IMACS]{Inamori-Magellan Areal Camera \& Spectrograph\acroextra{ (instrument on the Magellan Baade telescope)}}
\acro{IMR}[IMR]{inspiral-merger-ringdown}
\acro{IPAC}[IPAC]{Infrared Processing and Analysis Center}
\acro{IPN}[IPN]{InterPlanetary Network}
\acro{IPTF}[\acrolowercase{i}PTF]{intermediate \acl{PTF}}
\acro{ISM}[ISM]{interstellar medium}
\acro{ISS}[ISS]{International Space Station}
\acro{KAGRA}[KAGRA]{KAmioka GRAvitational\nobreakdashes-wave observatory}
\acro{KDE}[KDE]{kernel density estimator}
\acro{KN}[KN]{kilonova}
\acroplural{KN}[KNe]{kilonovae}
\acro{LAT}[LAT]{Large Area Telescope}
\acro{LCOGT}[LCOGT]{Las Cumbres Observatory Global Telescope}
\acro{LHO}[LHO]{\ac{LIGO} Hanford Observatory}
\acro{LIB}[LIB]{LALInference Burst}
\acro{LIGO}[LIGO]{Laser Interferometer \acs{GW} Observatory}
\acro{llGRB}[\acrolowercase{ll}GRB]{low\nobreakdashes-luminosity \ac{GRB}}
\acro{LLOID}[LLOID]{Low Latency Online Inspiral Detection}
\acro{LLO}[LLO]{\ac{LIGO} Livingston Observatory}
\acro{LMI}[LMI]{Large Monolithic Imager\acroextra{ (instrument on \ac{DCT})}}
\acro{LOFAR}[LOFAR]{Low Frequency Array}
\acro{LOS}[LOS]{line of sight}
\acroplural{LOS}[LOSs]{lines of sight}
\acro{LMC}[LMC]{Large Magellanic Cloud}
\acro{LSB}[LSB]{long, soft burst}
\acro{LSC}[LSC]{\acs{LIGO} Scientific Collaboration}
\acro{LSO}[LSO]{last stable orbit}
\acro{LSST}[LSST]{Large Synoptic Survey Telescope}
\acro{LT}[LT]{Liverpool Telescope}
\acro{LTI}[LTI]{linear time invariant}
\acro{MAP}[MAP]{maximum a posteriori}
\acro{MBTA}[MBTA]{Multi-Band Template Analysis}
\acro{MCMC}[MCMC]{Markov chain Monte Carlo}
\acro{MLE}[MLE]{\ac{ML} estimator}
\acro{ML}[ML]{maximum likelihood}
\acro{MOU}[MOU]{memorandum of understanding}
\acroplural{MOU}[MOUs]{memoranda of understanding}
\acro{MWA}[MWA]{Murchison Widefield Array}
\acro{NED}[NED]{NASA/IPAC Extragalactic Database}
\acro{NSBH}[NSBH]{neutron star\nobreakdashes--black hole}
\acro{NSBH}[NSBH]{\acl{NS}\nobreakdashes--\acl{BH}}
\acro{NSF}[NSF]{National Science Foundation}
\acro{NSNS}[NSNS]{\acl{NS}\nobreakdashes--\acl{NS}}
\acro{NS}[NS]{neutron star}
\acro{O1}[O1]{\acl{aLIGO}'s first observing run}
\acro{oLIB}[\acrolowercase{o}LIB]{Omicron+\acl{LIB}}
\acro{OT}[OT]{optical transient}
\acro{P48}[P48]{Palomar 48~inch Oschin telescope}
\acro{P60}[P60]{robotic Palomar 60~inch telescope}
\acro{P200}[P200]{Palomar 200~inch Hale telescope}
\acro{PC}[PC]{photon counting}
\acro{PESSTO}[PESSTO]{Public ESO Spectroscopic Survey of Transient Objects}
\acro{PSD}[PSD]{power spectral density}
\acro{PSF}[PSF]{point-spread function}
\acro{PS1}[PS1]{Pan\nobreakdashes-STARRS~1}
\acro{PTF}[PTF]{Palomar Transient Factory}
\acro{QUEST}[QUEST]{Quasar Equatorial Survey Team}
\acro{RAPTOR}[RAPTOR]{Rapid Telescopes for Optical Response}
\acro{REU}[REU]{Research Experiences for Undergraduates}
\acro{RMS}[RMS]{root mean square}
\acro{ROTSE}[ROTSE]{Robotic Optical Transient Search}
\acro{S5}[S5]{\ac{LIGO}'s fifth science run}
\acro{S6}[S6]{\ac{LIGO}'s sixth science run}
\acro{SAA}[SAA]{South Atlantic Anomaly}
\acro{SHB}[SHB]{short, hard burst}
\acro{SHGRB}[SHGRB]{short, hard \acl{GRB}}
\acro{SKA}[SKA]{Square Kilometer Array}
\acro{SMT}[SMT]{Slewing Mirror Telescope\acroextra{ (instrument on \acs{UFFO} Pathfinder)}}
\acro{SNR}[S/N]{signal\nobreakdashes-to\nobreakdashes-noise ratio}
\acro{SSC}[SSC]{synchrotron self\nobreakdashes-Compton}
\acro{SDSS}[SDSS]{Sloan Digital Sky Survey}
\acro{SED}[SED]{spectral energy distribution}
\acro{SGRB}[SGRB]{short \acl{GRB}}
\acro{SN}[SN]{supernova}
\acroplural{SN}[SN\acrolowercase{e}]{supernova}
\acro{SNIa}[\acs{SN}\,I\acrolowercase{a}]{Type~Ia \ac{SN}}
\acroplural{SNIa}[\acsp{SN}\,I\acrolowercase{a}]{Type~Ic \acp{SN}}
\acro{SNIcBL}[\acs{SN}\,I\acrolowercase{c}\nobreakdashes-BL]{broad\nobreakdashes-line Type~Ic \ac{SN}}
\acroplural{SNIcBL}[\acsp{SN}\,I\acrolowercase{c}\nobreakdashes-BL]{broad\nobreakdashes-line Type~Ic \acp{SN}}
\acro{SVD}[SVD]{singular value decomposition}
\acro{TAROT}[TAROT]{T\'{e}lescopes \`{a} Action Rapide pour les Objets Transitoires}
\acro{TDOA}[TDOA]{time delay on arrival}
\acroplural{TDOA}[TDOA\acrolowercase{s}]{time delays on arrival}
\acro{TD}[TD]{time domain}
\acro{TOA}[TOA]{time of arrival}
\acroplural{TOA}[TOA\acrolowercase{s}]{times of arrival}
\acro{TOO}[TOO]{target\nobreakdashes-of\nobreakdashes-opportunity}
\acroplural{TOO}[TOO\acrolowercase{s}]{targets of opportunity}
\acro{UFFO}[UFFO]{Ultra Fast Flash Observatory}
\acro{UHE}[UHE]{ultra high energy}
\acro{UVOT}[UVOT]{UV/Optical Telescope\acroextra{ (instrument on \emph{Swift})}}
\acro{VHE}[VHE]{very high energy}
\acro{VISTA}[VISTA@ESO]{Visible and Infrared Survey Telescope}
\acro{VLA}[VLA]{Karl G. Jansky Very Large Array}
\acro{VLT}[VLT]{Very Large Telescope}
\acro{VST}[VST@ESO]{\acs{VLT} Survey Telescope}
\acro{WAM}[WAM]{Wide\nobreakdashes-band All\nobreakdashes-sky Monitor\acroextra{ (instrument on \emph{Suzaku})}}
\acro{WCS}[WCS]{World Coordinate System}
\acro{WSS}[w.s.s.]{wide\nobreakdashes-sense stationary}
\acro{XRF}[XRF]{X\nobreakdashes-ray flash}
\acroplural{XRF}[XRF\acrolowercase{s}]{X\nobreakdashes-ray flashes}
\acro{XRT}[XRT]{X\nobreakdashes-ray Telescope\acroextra{ (instrument on \emph{Swift})}}
\acro{ZTF}[ZTF]{Zwicky Transient Facility}
\end{acronym}
 
\acknowledgements

\software{Astropy \citep{Astropy}, HEALPix \citep{HEALPix}}
 
 The authors gratefully acknowledge the support of the United States
National Science Foundation (NSF) for the construction and operation of the
LIGO Laboratory and Advanced LIGO as well as the Science and Technology Facilities Council (STFC) of the
United Kingdom, the Max-Planck Society (MPS), and the State of
Niedersachsen/Germany for support of the construction of Advanced LIGO 
and construction and operation of the GEO\,600 detector. 
Additional support for Advanced LIGO was provided by the Australian Research Council.
The authors gratefully acknowledge the Italian Istituto Nazionale di Fisica Nucleare (INFN),  
the French Centre National de la Recherche Scientifique (CNRS), and
the Foundation for Fundamental Research on Matter supported by the Netherlands Organisation for Scientific Research, 
for the construction and operation of the Virgo detector,
and the creation and support  of the EGO consortium. 
The authors also gratefully acknowledge research support from these agencies as well as by 
the Council of Scientific and Industrial Research of India, 
Department of Science and Technology, India,
Science \& Engineering Research Board (SERB), India,
Ministry of Human Resource Development, India,
the Spanish Ministerio de Econom\'ia y Competitividad,
the Conselleria d'Economia i Competitivitat and Conselleria d'Educaci\'o, Cultura i Universitats of the Govern de les Illes Balears,
the National Science Centre of Poland,
the European Commission,
the Royal Society, 
the Scottish Funding Council, 
the Scottish Universities Physics Alliance, 
the Hungarian Scientific Research Fund (OTKA),
the Lyon Institute of Origins (LIO),
the National Research Foundation of Korea,
Industry Canada and the Province of Ontario through the Ministry of Economic Development and Innovation, 
the National Science and Engineering Research Council Canada,
Canadian Institute for Advanced Research,
the Brazilian Ministry of Science, Technology, and Innovation,
Russian Foundation for Basic Research,
the Leverhulme Trust, 
the Research Corporation, 
Ministry of Science and Technology (MOST), Taiwan,
and
the Kavli Foundation.
The authors gratefully acknowledge the support of the NSF, STFC, MPS, INFN, CNRS, and the
State of Niedersachsen/Germany for provision of computational resources.
 \medskip

The Australian SKA Pathfinder is part of the Australia Telescope National Facility which is managed by CSIRO. The operation of ASKAP is funded by the Australian Government with support from the National Collaborative Research Infrastructure Strategy. Establishment of the Murchison Radio-astronomy Observatory was funded by the Australian Government and the Government of Western Australia. ASKAP uses advanced supercomputing resources at the Pawsey Supercomputing Centre. We acknowledge the Wajarri Yamatji people as the traditional owners of the Observatory site.
 \medskip

A.J.C.T. acknowledges support from the Junta de Andaluc\'ia (Project
P07-TIC-03094) and Univ. of Auckland and NIWA for installing of the
Spanish BOOTES-3 station in New Zealand, and support from the Spanish
Ministry Projects AYA2012-39727-C03-01 and 2015-71718R.
 \medskip

Funding for the DES Projects has been provided by the United States Department of Energy, the United States National Science Foundation, the Ministry of Science and Education of Spain, 
the Science and Technology Facilities Council of the United Kingdom, the Higher Education Funding Council for England, the National Center for Supercomputing 
Applications at the University of Illinois at Urbana-Champaign, the Kavli Institute of Cosmological Physics at the University of Chicago, 
the Center for Cosmology and Astro-Particle Physics at the Ohio State University,
the Mitchell Institute for Fundamental Physics and Astronomy at Texas A\&M University, Financiadora de Estudos e Projetos, 
Funda{\c c}{\~a}o Carlos Chagas Filho de Amparo {\`a} Pesquisa do Estado do Rio de Janeiro, Conselho Nacional de Desenvolvimento Cient{\'i}fico e Tecnol{\'o}gico and 
the Minist{\'e}rio da Ci{\^e}ncia, Tecnologia e Inova{\c c}{\~a}o, the Deutsche Forschungsgemeinschaft, and the Collaborating Institutions in the Dark Energy Survey. 

The Collaborating Institutions are Argonne National Laboratory, the University of California at Santa Cruz, the University of Cambridge, Centro de Investigaciones Energ{\'e}ticas, 
Medioambientales y Tecnol{\'o}gicas-Madrid, the University of Chicago, University College London, the DES-Brazil Consortium, the University of Edinburgh, 
the Eidgen{\"o}ssische Technische Hochschule (ETH) Z{\"u}rich, 
Fermi National Accelerator Laboratory, the University of Illinois at Urbana-Champaign, the Institut de Ci{\`e}ncies de l'Espai (IEEC/CSIC), 
the Institut de F{\'i}sica d'Altes Energies, Lawrence Berkeley National Laboratory, the Ludwig-Maximilians Universit{\"a}t M{\"u}nchen and the associated Excellence Cluster Universe, 
the University of Michigan, the National Optical Astronomy Observatory, the University of Nottingham, The Ohio State University, the University of Pennsylvania, the University of Portsmouth, 
SLAC National Accelerator Laboratory, Stanford University, the University of Sussex, and Texas A\&M University.

The DES data management system is supported by the National Science Foundation under Grant Number AST-1138766.
The DES participants from Spanish institutions are partially supported by MINECO under grants AYA2012-39559, ESP2013-48274, FPA2013-47986, and Centro de Excelencia Severo Ochoa SEV-2012-0234.
Research leading to these results has received funding from the European Research Council under the European Union's Seventh Framework Programme (FP7/2007-2013) including ERC grant agreements 
 240672, 291329, and 306478.
  \medskip

The \textit{Fermi} LAT Collaboration acknowledges support for LAT development, operation, and data analysis from NASA and DOE (United States), CEA/Irfu and IN2P3/CNRS (France), ASI and INFN (Italy), MEXT, KEK, and JAXA (Japan), and the K.A.~Wallenberg Foundation, the Swedish Research Council and the National Space Board (Sweden). Science analysis support in the operations phase from INAF (Italy) and CNES (France) is also gratefully acknowledged. The \textit{Fermi} GBM Collaboration acknowledges the support of NASA in the United States and DRL in Germany.
 \medskip

GRAWITA acknowledges the support of INAF for the project ``Gravitational Wave Astronomy with the first detections of adLIGO and adVIRGO experiments.''
 \medskip

This works exploited data by INTEGRAL, an ESA project with instruments
and science data center funded by ESA member states (especially the PI
countries: Denmark, France, Germany, Italy, Switzerland, Spain), and
with the participation of Russia and the USA.  The SPI ACS detector
system has been provided by MPE Garching/Germany. We acknowledge the
German INTEGRAL support through DLR grant 50 OG 1101.
 \medskip

IPN work is supported in the US under NASA Grant NNX15AU74G. \medskip

This work is partly based on observations obtained with the Samuel Oschin 48\,in Telescope and the 60\,in Telescope at the Palomar Observatory as part of the Intermediate Palomar Transient Factory (iPTF) project, a scientific collaboration among the California Institute of Technology, Los Alamos National Laboratory, the University of Wisconsin, Milwaukee, the Oskar Klein Center, the Weizmann Institute of Science, the TANGO Program of the University System of Taiwan, and the Kavli Institute for the Physics and Mathematics of the Universe. M.M.K. and Y.C. acknowledge funding from the National Science Foundation PIRE program grant 1545949. A.A.M. acknowledges support from the Hubble Fellowship HST\nobreakdashes-HF\nobreakdashes-51325.01. Part of the research was carried out at the Jet Propulsion Laboratory, California Institute of Technology, under a contract with NASA.
 \medskip

J-GEM is financially supported by KAKENHI Grant No. 24103003, 15H00774,
and 15H00788 of MEXT Japan, 15H02069 and 15H02075 of JSPS, and the
``Optical and Near-Infrared Astronomy Inter-University Cooperation
Program'' supported by MEXT.
 \medskip

The Liverpool Telescope is operated on the island of La Palma by Liverpool John Moores University in the Spanish Observatorio del Roque de los Muchachos of the Instituto de Astrofisica de Canarias with financial support from the UK Science and Technology Facilities Council.
 \medskip

LOFAR, the Low Frequency Array designed and constructed by ASTRON, has facilities in several countries, which are owned by various parties (each with their own funding sources), and that are collectively operated by the International LOFAR Telescope (ILT) foundation under a joint scientific policy. R.~Fender acknowledges support from ERC Advanced Investigator Grant 267697.
 \medskip

MASTER Global Robotic Net is supported in parts by Lomonosov Moscow
State University Development programm, Moscow Union OPTICA , Russian
Science Foundation 16-12-00085, RFBR15-02-07875, National Research
Foundation of South Africa.
 \medskip

We thank JAXA and RIKEN for providing MAXI data. The MAXI team is partially supported by KAKENHI Grant Nos. 24103002, 24540239, 24740186, and 23000004 of MEXT, Japan.
 \medskip

This work uses the Murchison Radio-astronomy Observatory, operated by
CSIRO. We acknowledge the Wajarri Yamatji people as the traditional owners
of the observatory site. Support for the operation of the MWA is provided
by the Australian Government Department of Industry and Science and
Department of Education (National Collaborative Research Infrastructure
Strategy: NCRIS), under a contract to Curtin University administered by
Astronomy Australia Limited. The MWA acknowledges the iVEC Petabyte Data
Store and the Initiative in Innovative Computing and the CUDA Center for
Excellence sponsored by NVIDIA at Harvard University.
 \medskip

Pan-STARRS is supported by the University of Hawaii and the National
Aeronautics and Space Administration's Planetary Defense Office under
grant No. NNX14AM74G. The Pan-STARRS-LIGO effort is in collaboration
with the LIGO Consortium and supported by Queen's University Belfast.
The Pan-STARRS1 Sky Surveys have been made possible through
contributions by the Institute for Astronomy, the University of
Hawaii, the Pan-STARRS Project Office, the Max Planck Society and its
participating institutes, the Max Planck Institute for Astronomy,
Heidelberg, and the Max Planck Institute for Extraterrestrial Physics,
Garching, The Johns Hopkins University, Durham University, the
University of Edinburgh, the Queen's University Belfast, the
Harvard-Smithsonian Center for Astrophysics, the Las Cumbres
Observatory Global Telescope Network Incorporated, the National
Central University of Taiwan, the Space Telescope Science Institute,
and the National Aeronautics and Space Administration under grant
No. NNX08AR22G issued through the Planetary Science Division of the
NASA Science Mission Directorate, the National Science Foundation
grant No. AST-1238877, the University of Maryland, Eotvos Lorand
University (ELTE), and the Los Alamos National Laboratory. This work
is based (in part) on observations collected at the European
Organisation for Astronomical Research in the Southern Hemisphere,
Chile as part of PESSTO, (the Public ESO Spectroscopic Survey for
Transient Objects Survey) ESO programs 188.D-3003, 191.D-0935.

Some of the data presented herein were obtained at the Palomar
Observatory, California Institute of Technology.

S.J.S. acknowledges funding from the European Research Council under the
European Union's Seventh Framework Programme (FP7/2007-2013)/ERC Grant
agreement No. [291222] and STFC grants ST/I001123/1 and
ST/L000709/1. M.F. is supported by the European Union FP7 programme
through ERC grant No. 320360.  K.M. acknowledges support from the
STFC through an Ernest Rutherford Fellowship.

F.O.E. acknowledges support from FONDECYT through postdoctoral grant 3140326.
 \medskip

Parts of this research were conducted by the Australian Research Council Centre of Excellence for All-sky Astrophysics (CAASTRO), through project No. CE110001020.
 \medskip

Funding for \emph{Swift} is provided by NASA in the US, by the UK Space Agency in the UK, and by the Agenzia Spaziale Italiana (ASI) in Italy.  This work made use of data supplied by the UK \emph{Swift} Science Data Centre at the University of Leicester.  We acknowledge the use of public data from the \emph{Swift} data archive. \medskip

The TOROS collaboration acknowledges support from Ministerio de Ciencia y Tecnolog\'ia (MinCyT) and Consejo Nacional de Investigaciones Cientificas y Tecnol\'ogicas (CONICET) from Argentina and grants from the USA NSF PHYS 1156600 and NSF HRD 1242090.
 \medskip

The National Radio Astronomy Observatory is a facility of the National Science Foundation operated under cooperative agreement by Associated Universities, Inc.
 \medskip

VST and VISTA observations were performed at the European Southern Observatory, Paranal, Chile. We acknowledge ESO personnel for their assistance during the observing runs.
 \medskip

This is LIGO document LIGO-P1600137-v2.

\bibliographystyle{aasjournal}

\onecolumngrid
\centering
\section*{Authors}
\textsc{B.~P.~Abbott\textsuperscript{1}},
\textsc{R.~Abbott\textsuperscript{1}},
\textsc{T.~D.~Abbott\textsuperscript{2}},
\textsc{M.~R.~Abernathy\textsuperscript{1}},
\textsc{F.~Acernese\textsuperscript{3,4}},
\textsc{K.~Ackley\textsuperscript{5}},
\textsc{C.~Adams\textsuperscript{6}},
\textsc{T.~Adams\textsuperscript{7}},
\textsc{P.~Addesso\textsuperscript{3}},
\textsc{R.~X.~Adhikari\textsuperscript{1}}
% ...

(\textsc{The LIGO Scientific Collaboration and the Virgo Collaboration})
\vspace{1em}

\textsc{J.~Allison\textsuperscript{134,135}},
\textsc{K.~Bannister\textsuperscript{134,135}},
\textsc{M.~E.~Bell\textsuperscript{134,135}},
\textsc{S.~Chatterjee\textsuperscript{136}},
\textsc{A.~P.~Chippendale\textsuperscript{134}},
\textsc{P.~G.~Edwards\textsuperscript{134}},
\textsc{L.~Harvey-Smith\textsuperscript{134}},
\textsc{Ian~Heywood\textsuperscript{134,137}},
\textsc{A.~Hotan\textsuperscript{138}},
\textsc{B.~Indermuehle\textsuperscript{134}},
\textsc{J.~Marvil\textsuperscript{134}},
\textsc{D.~McConnell\textsuperscript{134}},
\textsc{T.~Murphy\textsuperscript{139,135}},
\textsc{A.~Popping\textsuperscript{140,135}},
\textsc{J.~Reynolds\textsuperscript{134}},
\textsc{R.~J.~Sault\textsuperscript{84,134}},
\textsc{M.~A.~Voronkov\textsuperscript{134}},
\textsc{M.~T.~Whiting\textsuperscript{134}}

(\textsc{The Australian Square Kilometer Array Pathfinder (ASKAP) Collaboration})
\vspace{1em}

\textsc{A.~J.~Castro-Tirado\textsuperscript{141,142}},
\textsc{R.~Cunniffe\textsuperscript{141}},
\textsc{M.~Jel\'inek\textsuperscript{143}},
\textsc{J.~C.~Tello\textsuperscript{141}},
\textsc{S.~R.~Oates\textsuperscript{141}},
\textsc{Y.-D.~Hu\textsuperscript{141}},
\textsc{P.~Kub\'anek\textsuperscript{144}},
\textsc{S.~Guziy\textsuperscript{145}},
\textsc{A.~Castell\'on\textsuperscript{146}},
\textsc{A.~Garc\'ia-Cerezo\textsuperscript{142}},
\textsc{V.~F.~Mu\~noz\textsuperscript{142}},
\textsc{C.~{P\'erez}~del~Pulgar\textsuperscript{142}},
\textsc{S.~Castillo-Carri\'on\textsuperscript{147}},
\textsc{J.~M.~{Castro}~Cer\'on\textsuperscript{148}},
\textsc{R.~Hudec\textsuperscript{143,149}},
\textsc{M.~D.~Caballero-Garc\'ia\textsuperscript{150}},
\textsc{P.~P\'ata\textsuperscript{149}},
\textsc{S.~Vitek\textsuperscript{149}},
\textsc{J.~A.~Adame\textsuperscript{151}},
\textsc{S.~Konig\textsuperscript{151}},
\textsc{F.~Rend\'on\textsuperscript{141,151}},
\textsc{T.~de~J.~{Mateo}~Sanguino\textsuperscript{152}},
\textsc{R.~Fern\'andez-Mu\~noz\textsuperscript{153}},
\textsc{P.~C.~Yock\textsuperscript{154}},
\textsc{N.~Rattenbury\textsuperscript{154}},
\textsc{W.~H.~Allen\textsuperscript{155}},
\textsc{R.~Querel\textsuperscript{156}},
\textsc{S.~Jeong\textsuperscript{141,157}},
\textsc{I.~H.~Park\textsuperscript{157}},
\textsc{J.~Bai\textsuperscript{158}},
\textsc{Ch.~Cui\textsuperscript{159}},
\textsc{Y.~Fan\textsuperscript{158}},
\textsc{Ch.~Wang\textsuperscript{158}},
\textsc{D.~Hiriart\textsuperscript{160}},
\textsc{W.~H.~Lee\textsuperscript{161}},
\textsc{A.~Claret\textsuperscript{141}},
\textsc{R.~S\'anchez-Ram\'irez\textsuperscript{141}},
\textsc{S.~B.~Pandey\textsuperscript{162}},
\textsc{T.~Mediavilla\textsuperscript{163}},
\textsc{L.~Sabau-Graziati\textsuperscript{164}}

(\textsc{The BOOTES Collaboration})
\vspace{1em}

\textsc{T.~M.~C.~Abbott\textsuperscript{165}},
\textsc{F.~B.~Abdalla\textsuperscript{166,137}},
\textsc{S.~Allam\textsuperscript{167}},
\textsc{J.~Annis\textsuperscript{167}},
\textsc{R.~Armstrong\textsuperscript{168}},
\textsc{A.~Benoit-L{\'e}vy\textsuperscript{169,166,170}},
\textsc{E.~Berger\textsuperscript{171}},
\textsc{R.~A.~Bernstein\textsuperscript{172}},
\textsc{E.~Bertin\textsuperscript{169,170}},
\textsc{D.~Brout\textsuperscript{173}
% ...
}

(\textsc{The Dark Energy Survey and the Dark Energy Camera GW-EM Collaborations})
\vspace{1em}

\textsc{V.~Connaughton\textsuperscript{213}},
\textsc{E.~Burns\textsuperscript{214}},
\textsc{A.~Goldstein\textsuperscript{215,216}},
\textsc{M.~S.~Briggs\textsuperscript{217}},
\textsc{B.-B.~Zhang\textsuperscript{218,219}},
\textsc{C.~M.~Hui\textsuperscript{216}},
\textsc{P.~Jenke\textsuperscript{218}},
\textsc{C.~A.~Wilson-Hodge\textsuperscript{216}},
\textsc{P.~N.~Bhat\textsuperscript{218}},
\textsc{E.~Bissaldi\textsuperscript{220}},
\textsc{W.~Cleveland\textsuperscript{213}},
\textsc{G.~Fitzpatrick\textsuperscript{218}},
\textsc{M.~M.~Giles\textsuperscript{221}},
\textsc{M.~H.~Gibby\textsuperscript{221}},
\textsc{J.~Greiner\textsuperscript{222}},
\textsc{A.~von~Kienlin\textsuperscript{222}},
\textsc{R.~M.~Kippen\textsuperscript{223}},
\textsc{S.~McBreen\textsuperscript{224}},
\textsc{B.~Mailyan\textsuperscript{218}},
\textsc{C.~A.~Meegan\textsuperscript{218}},
\textsc{W.~S.~Paciesas\textsuperscript{213}},
\textsc{R.~D.~Preece\textsuperscript{217}},
\textsc{O.~Roberts\textsuperscript{224}},
\textsc{L.~Sparke\textsuperscript{225}},
\textsc{M.~Stanbro\textsuperscript{214}},
\textsc{K.~Toelge\textsuperscript{222}},
\textsc{P.~Veres\textsuperscript{218}},
\textsc{H.-F.~Yu\textsuperscript{222,226}},
\textsc{L.~Blackburn\textsuperscript{171}}

(\textsc{The \emph{Fermi} GBM Collaboration})
\vspace{1em}

\textsc{M.~Ackermann\textsuperscript{227}},
\textsc{M.~Ajello\textsuperscript{228}},
\textsc{A.~Albert\textsuperscript{229}},
\textsc{B.~Anderson\textsuperscript{230,231}},
\textsc{W.~B.~Atwood\textsuperscript{232}},
\textsc{M.~Axelsson\textsuperscript{233,234}},
\textsc{L.~Baldini\textsuperscript{235,229}},
\textsc{G.~Barbiellini\textsuperscript{236,237}},
\textsc{D.~Bastieri\textsuperscript{238,239}},
\textsc{R.~Bellazzini\textsuperscript{240}}
% ...

(\textsc{The \emph{Fermi} LAT Collaboration})
\vspace{1em}

\textsc{E.~Brocato\textsuperscript{285}},
\textsc{E.~Cappellaro\textsuperscript{286}},
\textsc{S.~Covino\textsuperscript{287}},
\textsc{A.~Grado\textsuperscript{288}},
\textsc{L.~Nicastro\textsuperscript{289}},
\textsc{E.~Palazzi\textsuperscript{289}},
\textsc{E.~Pian\textsuperscript{289,290}},
\textsc{L.~Amati\textsuperscript{289}},
\textsc{L.~A.~Antonelli\textsuperscript{285,291}},
\textsc{M.~Capaccioli\textsuperscript{292}},
\textsc{P.~D'Avanzo\textsuperscript{287}},
\textsc{V.~D'Elia\textsuperscript{285,291}},
\textsc{F.~Getman\textsuperscript{288}},
\textsc{G.~Giuffrida\textsuperscript{285,291}},
\textsc{G.~Iannicola\textsuperscript{285}},
\textsc{L.~Limatola\textsuperscript{288}},
\textsc{M.~Lisi\textsuperscript{285}},
\textsc{S.~Marinoni\textsuperscript{285,291}},
\textsc{P.~Marrese\textsuperscript{285,286}},
\textsc{A.~Melandri\textsuperscript{287}},
\textsc{S.~Piranomonte\textsuperscript{285}},
\textsc{A.~Possenti\textsuperscript{293}},
\textsc{L.~Pulone\textsuperscript{285}},
\textsc{A.~Rossi\textsuperscript{289}},
\textsc{A.~Stamerra\textsuperscript{290,294}},
\textsc{L.~Stella\textsuperscript{285}},
\textsc{V.~Testa\textsuperscript{285}},
\textsc{L.~Tomasella\textsuperscript{286}},
\textsc{S.~Yang\textsuperscript{286}}

(\textsc{The GRAvitational Wave Inaf TeAm (GRAWITA)})
\vspace{1em}

\textsc{A.~Bazzano\textsuperscript{295}},
\textsc{E.~Bozzo\textsuperscript{296}},
\textsc{S.~Brandt\textsuperscript{297}},
\textsc{T.~J.-L.~Courvoisier\textsuperscript{296}},
\textsc{C.~Ferrigno\textsuperscript{296}},
\textsc{L.~Hanlon\textsuperscript{298}},
\textsc{E.~Kuulkers\textsuperscript{299}},
\textsc{P.~Laurent\textsuperscript{300}},
\textsc{S.~Mereghetti\textsuperscript{301}},
\textsc{J.~P.~Roques\textsuperscript{302}},
\textsc{V.~Savchenko\textsuperscript{303}},
\textsc{P.~Ubertini\textsuperscript{295}}

(\textsc{The \emph{INTEGRAL} Collaboration})
\vspace{1em}

\textsc{M.~M.~Kasliwal\textsuperscript{304}},
\textsc{L.~P.~Singer\textsuperscript{39,215}},
\textsc{Y.~Cao\textsuperscript{304}},
\textsc{G.~Duggan\textsuperscript{304}},
\textsc{S.~R.~Kulkarni\textsuperscript{304}},
\textsc{V.~Bhalerao\textsuperscript{14}},
\textsc{A.~A.~Miller\textsuperscript{185,304,305}},
\textsc{T.~Barlow\textsuperscript{304}},
\textsc{E.~Bellm\textsuperscript{304}},
\textsc{I.~Manulis\textsuperscript{306}},
\textsc{J.~Rana\textsuperscript{14}},
\textsc{R.~Laher\textsuperscript{307}},
\textsc{F.~Masci\textsuperscript{307}},
\textsc{J.~Surace\textsuperscript{307}},
\textsc{U.~Rebbapragada\textsuperscript{185}},
\textsc{D.~Cook\textsuperscript{304}},
\textsc{A.~Van~Sistine\textsuperscript{16}},
\textsc{B.~Sesar\textsuperscript{308}},
\textsc{D.~Perley\textsuperscript{309}},
\textsc{R.~Ferreti\textsuperscript{310}},
\textsc{T.~Prince\textsuperscript{304}},
\textsc{R.~Kendrick\textsuperscript{311}},
\textsc{A.~Horesh\textsuperscript{306}}

(\textsc{The Intermediate Palomar Transient Factory (iPTF) Collaboration})
\vspace{1em}

\textsc{K.~Hurley\textsuperscript{312}},
\textsc{S.~V.~Golenetskii\textsuperscript{313}},
\textsc{R.~L.~Aptekar\textsuperscript{313}},
\textsc{D.~D.~Frederiks\textsuperscript{313}},
\textsc{D.~S.~Svinkin\textsuperscript{313}},
\textsc{A.~Rau\textsuperscript{222}},
\textsc{A.~von~Kienlin\textsuperscript{222}},
\textsc{X.~Zhang\textsuperscript{222}},
\textsc{D.~M.~Smith\textsuperscript{232}},
\textsc{T.~Cline\textsuperscript{39,314}},
\textsc{H.~Krimm\textsuperscript{247,315}}

(\textsc{The InterPlanetary Network})
\vspace{1em}

\textsc{F.~Abe\textsuperscript{316}},
\textsc{M.~Doi\textsuperscript{317}},
\textsc{K.~Fujisawa\textsuperscript{318}},
\textsc{K.~S.~Kawabata\textsuperscript{319}},
\textsc{T.~Morokuma\textsuperscript{317}},
\textsc{K.~Motohara\textsuperscript{317}},
\textsc{M.~Tanaka\textsuperscript{320}},
\textsc{K.~Ohta\textsuperscript{321}},
\textsc{K.~Yanagisawa\textsuperscript{322}},
\textsc{M.~Yoshida\textsuperscript{319}}

(\textsc{The J-GEM Collaboration})
\vspace{1em}

\textsc{C.~Baltay\textsuperscript{323}},
\textsc{D.~Rabinowitz\textsuperscript{323}},
\textsc{N.~Ellman\textsuperscript{323}},
\textsc{S.~Rostami\textsuperscript{323}}

(\textsc{The La Silla--QUEST Survey})
\vspace{1em}

\textsc{D.~F.~Bersier\textsuperscript{324}},
\textsc{M.~F.~Bode\textsuperscript{324}},
\textsc{C.~A.~Collins\textsuperscript{324}},
\textsc{C.~M.~Copperwheat\textsuperscript{324}},
\textsc{M.~J.~Darnley\textsuperscript{324}},
\textsc{D.~K.~Galloway\textsuperscript{325,326}},
\textsc{A.~Gomboc\textsuperscript{327,328}},
\textsc{S.~Kobayashi\textsuperscript{324}},
\textsc{P.~Mazzali\textsuperscript{324}},
\textsc{C.~G.~Mundell\textsuperscript{329}},
\textsc{A.~S.~Piascik\textsuperscript{324}},
\textsc{Don~Pollacco\textsuperscript{330}},
\textsc{I.~A.~Steele\textsuperscript{324}},
\textsc{K.~Ulaczyk\textsuperscript{330}}

(\textsc{The Liverpool Telescope Collaboration})
\vspace{1em}

\textsc{J.W.~Broderick\textsuperscript{331}},
\textsc{R.P.~Fender\textsuperscript{332}},
\textsc{P.G.~Jonker\textsuperscript{333,52}},
\textsc{A.~Rowlinson\textsuperscript{331,334,135}},
\textsc{B.W.~Stappers\textsuperscript{212}},
\textsc{R.A.M.J.~Wijers\textsuperscript{334}}

(\textsc{The Low Frequency Array (LOFAR) Collaboration})
\vspace{1em}

\textsc{V.~Lipunov\textsuperscript{335}},
\textsc{E.~Gorbovskoy\textsuperscript{335}},
\textsc{N.~Tyurina\textsuperscript{335}},
\textsc{V.~Kornilov\textsuperscript{335}},
\textsc{P.~Balanutsa\textsuperscript{335}},
\textsc{A.~Kuznetsov\textsuperscript{335}},
\textsc{D.~Buckley\textsuperscript{336}},
\textsc{R.~Rebolo\textsuperscript{337}},
\textsc{M.~Serra-Ricart\textsuperscript{337}},
\textsc{G.~Israelian\textsuperscript{337}},
\textsc{N.~M.~Budnev\textsuperscript{338}},
\textsc{O.~Gress\textsuperscript{338}},
\textsc{K.~Ivanov\textsuperscript{338}},
\textsc{V.~Poleshuk\textsuperscript{338}},
\textsc{A.~Tlatov\textsuperscript{339}},
\textsc{V.~Yurkov\textsuperscript{340}}

(\textsc{The MASTER Collaboration})
\vspace{1em}

\textsc{N.~Kawai\textsuperscript{341}},
\textsc{M.~Serino\textsuperscript{342}},
\textsc{H.~Negoro\textsuperscript{343}},
\textsc{S.~Nakahira\textsuperscript{344}},
\textsc{T.~Mihara\textsuperscript{342}},
\textsc{H.~Tomida\textsuperscript{345}},
\textsc{S.~Ueno\textsuperscript{345}},
\textsc{H.~Tsunemi\textsuperscript{346}},
\textsc{M.~Matsuoka\textsuperscript{342}}

(\textsc{The MAXI Collaboration})
\vspace{1em}

\textsc{S.~Croft\textsuperscript{347,348}},
\textsc{L.~Feng\textsuperscript{349}},
\textsc{T.~M.~O.~Franzen\textsuperscript{350}},
\textsc{B.~M.~Gaensler\textsuperscript{351,139,135}},
\textsc{M.~Johnston-Hollitt\textsuperscript{352}},
\textsc{D.~L.~Kaplan\textsuperscript{16}},
\textsc{M.~F.~Morales\textsuperscript{131}},
\textsc{S.~J.~Tingay\textsuperscript{350,135,353}},
\textsc{R.~B.~Wayth\textsuperscript{350,135}},
\textsc{A.~Williams\textsuperscript{350}}

(\textsc{The Murchison Wide-field Array (MWA) Collaboration})
\vspace{1em}

\textsc{S.~J.~Smartt\textsuperscript{354}},
\textsc{K.~C.~Chambers\textsuperscript{355}},
\textsc{K.~W.~Smith\textsuperscript{354}},
\textsc{M.~E.~Huber\textsuperscript{355}},
\textsc{D.~R.~Young\textsuperscript{354}},
\textsc{D.~E.~Wright\textsuperscript{354}},
\textsc{A.~Schultz\textsuperscript{355}},
\textsc{L.~Denneau\textsuperscript{355}},
\textsc{H.~Flewelling\textsuperscript{355}},
\textsc{E.~A.~Magnier\textsuperscript{355}},
\textsc{N.~Primak\textsuperscript{355}},
\textsc{A.~Rest\textsuperscript{356}},
\textsc{A.~Sherstyuk\textsuperscript{355}},
\textsc{B.~Stalder\textsuperscript{355}},
\textsc{C.~W.~Stubbs\textsuperscript{357}},
\textsc{J.~Tonry\textsuperscript{355}},
\textsc{C.~Waters\textsuperscript{355}},
\textsc{M.~Willman\textsuperscript{355}}

(\textsc{The Pan-STARRS Collaboration})
\vspace{1em}

\textsc{F.~Olivares~E.\textsuperscript{358,359}},
\textsc{H.~Campbell\textsuperscript{360}},
\textsc{R.~Kotak\textsuperscript{354}},
\textsc{J.~Sollerman\textsuperscript{310}},
\textsc{M.~Smith\textsuperscript{26}},
\textsc{M.~Dennefeld\textsuperscript{361}},
\textsc{J.~P.~Anderson\textsuperscript{362}},
\textsc{M.~T.~Botticella\textsuperscript{288}},
\textsc{T.-W.~Chen\textsuperscript{222}},
\textsc{M.~D.~Valle\textsuperscript{288}},
\textsc{N.~Elias-Rosa\textsuperscript{286}},
\textsc{M.~Fraser\textsuperscript{360}},
\textsc{C.~Inserra\textsuperscript{354}},
\textsc{E.~Kankare\textsuperscript{354}},
\textsc{T.~Kupfer\textsuperscript{304}},
\textsc{J.~Harmanen\textsuperscript{363}},
\textsc{L.~Galbany\textsuperscript{358,364}},
\textsc{L.~Le~Guillou\textsuperscript{365,366}},
\textsc{J.~D.~Lyman\textsuperscript{330}},
\textsc{K.~Maguire\textsuperscript{354}},
\textsc{A.~Mitra\textsuperscript{366}},
\textsc{M.~Nicholl\textsuperscript{171}},
\textsc{A.~Razza\textsuperscript{358,364}},
\textsc{G.~Terreran\textsuperscript{286,354}},
\textsc{S.~Valenti\textsuperscript{367,368}},
\textsc{A.~Gal-Yam\textsuperscript{369}}

(\textsc{The PESSTO Collaboration})
\vspace{1em}

\textsc{A.~\'Cwiek\textsuperscript{110}},
\textsc{M.~\'Cwiok\textsuperscript{370}},
\textsc{L.~Mankiewicz\textsuperscript{371}},
\textsc{R.~Opiela\textsuperscript{371}},
\textsc{M.~Zaremba\textsuperscript{370}},
\textsc{A.~F.~\.Zarnecki\textsuperscript{370}}

(\textsc{The Pi of the Sky Collaboration})
\vspace{1em}

\textsc{C.~A.~Onken\textsuperscript{20,135}},
\textsc{R.~A.~Scalzo\textsuperscript{20,135}},
\textsc{B.~P.~Schmidt\textsuperscript{20,135}},
\textsc{C.~Wolf\textsuperscript{20,135}},
\textsc{F.~Yuan\textsuperscript{20,135}}

(\textsc{The SkyMapper Collaboration})
\vspace{1em}

\textsc{P.~A.~Evans\textsuperscript{372}},
\textsc{J.~A.~Kennea\textsuperscript{72}},
\textsc{D.~N.~Burrows\textsuperscript{72}},
\textsc{S.~Campana\textsuperscript{287}},
\textsc{S.~B.~Cenko\textsuperscript{39,373}},
\textsc{P.~Giommi\textsuperscript{291}},
\textsc{F.~E.~Marshall\textsuperscript{39}},
\textsc{J.~Nousek\textsuperscript{72}},
\textsc{P.~O'Brien\textsuperscript{372}},
\textsc{J.~P.~Osborne\textsuperscript{372}},
\textsc{D.~Palmer\textsuperscript{374}},
\textsc{M.~Perri\textsuperscript{291,285}},
\textsc{M.~Siegel\textsuperscript{72}},
\textsc{G.~Tagliaferri\textsuperscript{287}}

(\textsc{The \emph{Swift} Collaboration})
\vspace{1em}

\textsc{A.~Klotz\textsuperscript{375}},
\textsc{D.~Turpin\textsuperscript{375}},
\textsc{R.~Laugier\textsuperscript{53}}

(\textsc{The TAROT, Zadko, Algerian National Observatory, and C2PU Collaboration})
\vspace{1em}

\textsc{M.~Beroiz\textsuperscript{85,376}},
\textsc{T.~Pe\~{n}uela\textsuperscript{85,183}},
\textsc{L.~M.~Macri\textsuperscript{377}},
\textsc{R.~J.~Oelkers\textsuperscript{377}},
\textsc{D.~G.~Lambas\textsuperscript{378}},
\textsc{R.~Vrech\textsuperscript{378}},
\textsc{J.~Cabral\textsuperscript{378}},
\textsc{C.~Colazo\textsuperscript{378}},
\textsc{M.~Dominguez\textsuperscript{378}},
\textsc{B.~Sanchez\textsuperscript{378}},
\textsc{S.~Gurovich\textsuperscript{378}},
\textsc{M.~Lares\textsuperscript{378}},
\textsc{J.~L.~Marshall\textsuperscript{377}},
\textsc{D.~L.~DePoy\textsuperscript{377}},
\textsc{N.~Padilla\textsuperscript{379}},
\textsc{N.~A.~Pereyra\textsuperscript{85}},
\textsc{M.~Benacquista\textsuperscript{85}}

(\textsc{The TOROS Collaboration})
\vspace{1em}

\textsc{N.~R.~Tanvir\textsuperscript{372}},
\textsc{K.~Wiersema\textsuperscript{372}},
\textsc{A.~J.~Levan\textsuperscript{330}},
\textsc{D.~Steeghs\textsuperscript{330}},
\textsc{J.~Hjorth\textsuperscript{309}},
\textsc{J.~P.~U.~Fynbo\textsuperscript{309}},
\textsc{D.~Malesani\textsuperscript{309}},
\textsc{B.~Milvang-Jensen\textsuperscript{309}},
\textsc{D.~Watson\textsuperscript{309}},
\textsc{M.~Irwin\textsuperscript{360}},
\textsc{C.~G.~Fernandez\textsuperscript{360}},
\textsc{R.~G.~McMahon\textsuperscript{360}},
\textsc{M.~Banerji\textsuperscript{360}},
\textsc{E.~Gonzalez-Solares\textsuperscript{360}},
\textsc{S.~Schulze\textsuperscript{379,358}},
\textsc{A.~de~U.~Postigo\textsuperscript{380,309}},
\textsc{C.~C.~Thoene\textsuperscript{380}},
\textsc{Z.~Cano\textsuperscript{381}},
\textsc{S.~Rosswog\textsuperscript{310}}

(\textsc{The VISTA Collaboration})
\vspace{1em}

\textsuperscript{1}{\footnotesize{LIGO, California Institute of Technology, Pasadena, CA 91125, USA}}\\
\textsuperscript{2}{\footnotesize{Louisiana State University, Baton Rouge, LA 70803, USA}}\\
\textsuperscript{3}{\footnotesize{Universit\`a di Salerno, Fisciano, I-84084 Salerno, Italy}}\\
\textsuperscript{4}{\footnotesize{INFN, Sezione di Napoli, Complesso Universitario di Monte S.Angelo, I-80126 Napoli, Italy}}\\
\textsuperscript{5}{\footnotesize{University of Florida, Gainesville, FL 32611, USA}}\\
\textsuperscript{6}{\footnotesize{LIGO Livingston Observatory, Livingston, LA 70754, USA}}\\
\textsuperscript{7}{\footnotesize{Laboratoire d'Annecy-le-Vieux de Physique des Particules (LAPP), Universit\'e Savoie Mont Blanc, CNRS/IN2P3, F-74941 Annecy-le-Vieux, France}}\\
\textsuperscript{8}{\footnotesize{Albert-Einstein-Institut, Max-Planck-Institut f\"ur Gravi\-ta\-tions\-physik, D-30167 Hannover, Germany}}\\
\textsuperscript{9}{\footnotesize{Nikhef, Science Park, 1098 XG Amsterdam, The Netherlands}}\\
\textsuperscript{10}{\footnotesize{LIGO, Massachusetts Institute of Technology, Cambridge, MA 02139, USA}}\\
\textsuperscript{11}{\footnotesize{Instituto Nacional de Pesquisas Espaciais, 12227-010 S\~{a}o Jos\'{e} dos Campos, SP, Brazil}}\\
\textsuperscript{12}{\footnotesize{INFN, Gran Sasso Science Institute, I-67100 L'Aquila, Italy}}\\
\textsuperscript{13}{\footnotesize{{Istituto Nazionale di Fisica Nucleare, Sezione di Roma ``Tor Vergata," I-00133 Roma, Italy}}}\\
\textsuperscript{14}{\footnotesize{Inter-University Centre for Astronomy and Astrophysics, Pune 411007, India}}\\
\textsuperscript{15}{\footnotesize{International Centre for Theoretical Sciences, Tata Institute of Fundamental Research, Bangalore 560012, India}}\\
\textsuperscript{16}{\footnotesize{University of Wisconsin-Milwaukee, Milwaukee, WI 53201, USA}}\\
\textsuperscript{17}{\footnotesize{Leibniz Universit\"at Hannover, D-30167 Hannover, Germany}}\\
\textsuperscript{18}{\footnotesize{Universit\`a di Pisa, I-56127 Pisa, Italy}}\\
\textsuperscript{19}{\footnotesize{INFN, Sezione di Pisa, I-56127 Pisa, Italy}}\\
\textsuperscript{20}{\footnotesize{Australian National University, Canberra, Australian Capital Territory 0200, Australia}}\\
\textsuperscript{21}{\footnotesize{The University of Mississippi, University, MS 38677, USA}}\\
\textsuperscript{22}{\footnotesize{California State University Fullerton, Fullerton, CA 92831, USA}}\\
\textsuperscript{23}{\footnotesize{LAL, Univ. Paris-Sud, CNRS/IN2P3, Universit\'e Paris-Saclay, Orsay, France}}\\
\textsuperscript{24}{\footnotesize{Chennai Mathematical Institute, Chennai, India}}\\
\textsuperscript{25}{\footnotesize{Universit\`a di Roma ``Tor Vergata,'' I-00133 Roma, Italy}}\\
\textsuperscript{26}{\footnotesize{School of Physics and Astronomy, University of Southampton,  Southampton, SO17 1BJ, UK}}\\
\textsuperscript{27}{\footnotesize{Universit\"at Hamburg, D-22761 Hamburg, Germany}}\\
\textsuperscript{28}{\footnotesize{INFN, Sezione di Roma, I-00185 Roma, Italy}}\\
\textsuperscript{29}{\footnotesize{Albert-Einstein-Institut, Max-Planck-Institut f\"ur Gravitations\-physik, D-14476 Potsdam-Golm, Germany}}\\
\textsuperscript{30}{\footnotesize{APC, AstroParticule et Cosmologie, Universit\'e Paris Diderot, CNRS/IN2P3, CEA/Irfu, Observatoire de Paris, Sorbonne Paris Cit\'e, F-75205 Paris Cedex 13, France}}\\
\textsuperscript{31}{\footnotesize{Montana State University, Bozeman, MT 59717, USA}}\\
\textsuperscript{32}{\footnotesize{Universit\`a di Perugia, I-06123 Perugia, Italy}}\\
\textsuperscript{33}{\footnotesize{INFN, Sezione di Perugia, I-06123 Perugia, Italy}}\\
\textsuperscript{34}{\footnotesize{European Gravitational Observatory (EGO), I-56021 Cascina, Pisa, Italy}}\\
\textsuperscript{35}{\footnotesize{Syracuse University, Syracuse, NY 13244, USA}}\\
\textsuperscript{36}{\footnotesize{SUPA, University of Glasgow, Glasgow G12 8QQ, United Kingdom}}\\
\textsuperscript{37}{\footnotesize{LIGO Hanford Observatory, Richland, WA 99352, USA}}\\
\textsuperscript{38}{\footnotesize{Wigner RCP, RMKI, H-1121 Budapest, Konkoly Thege Mikl\'os \'ut 29-33, Hungary}}\\
\textsuperscript{39}{\footnotesize{NASA/Goddard Space Flight Center, Greenbelt, MD 20771, USA}}\\
\textsuperscript{40}{\footnotesize{Columbia University, New York, NY 10027, USA}}\\
\textsuperscript{41}{\footnotesize{Stanford University, Stanford, CA 94305, USA}}\\
\textsuperscript{42}{\footnotesize{Universit\`a di Padova, Dipartimento di Fisica e Astronomia, I-35131 Padova, Italy}}\\
\textsuperscript{43}{\footnotesize{INFN, Sezione di Padova, I-35131 Padova, Italy}}\\
\textsuperscript{44}{\footnotesize{CAMK-PAN, 00-716 Warsaw, Poland}}\\
\textsuperscript{45}{\footnotesize{University of Birmingham, Birmingham B15 2TT, United Kingdom}}\\
\textsuperscript{46}{\footnotesize{Universit\`a degli Studi di Genova, I-16146 Genova, Italy}}\\
\textsuperscript{47}{\footnotesize{INFN, Sezione di Genova, I-16146 Genova, Italy}}\\
\textsuperscript{48}{\footnotesize{RRCAT, Indore MP 452013, India}}\\
\textsuperscript{49}{\footnotesize{Faculty of Physics, Lomonosov Moscow State University, Moscow 119991, Russia}}\\
\textsuperscript{50}{\footnotesize{SUPA, University of the West of Scotland, Paisley PA1 2BE, United Kingdom}}\\
\textsuperscript{51}{\footnotesize{University of Western Australia, Crawley, Western Australia 6009, Australia}}\\
\textsuperscript{52}{\footnotesize{Department of Astrophysics/IMAPP, Radboud University Nijmegen, P.O. Box 9010, 6500 GL Nijmegen, The Netherlands}}\\
\textsuperscript{53}{\footnotesize{Artemis, Universit\'e C\^ote d'Azur, CNRS, Observatoire C\^ote d'Azur, CS 34229, Nice cedex 4, France}}\\
\textsuperscript{54}{\footnotesize{MTA E\"otv\"os University, ``Lendulet'' Astrophysics Research Group, Budapest 1117, Hungary}}\\
\textsuperscript{55}{\footnotesize{Institut de Physique de Rennes, CNRS, Universit\'e de Rennes 1, F-35042 Rennes, France}}\\
\textsuperscript{56}{\footnotesize{Washington State University, Pullman, WA 99164, USA}}\\
\textsuperscript{57}{\footnotesize{Universit\`a degli Studi di Urbino ``Carlo Bo,'' I-61029 Urbino, Italy}}\\
\textsuperscript{58}{\footnotesize{INFN, Sezione di Firenze, I-50019 Sesto Fiorentino, Firenze, Italy}}\\
\textsuperscript{59}{\footnotesize{University of Oregon, Eugene, OR 97403, USA}}\\
\textsuperscript{60}{\footnotesize{Laboratoire Kastler Brossel, UPMC-Sorbonne Universit\'es, CNRS, ENS-PSL Research University, Coll\`ege de France, F-75005 Paris, France}}\\
\textsuperscript{61}{\footnotesize{Astronomical Observatory Warsaw University, 00-478 Warsaw, Poland}}\\
\textsuperscript{62}{\footnotesize{VU University Amsterdam, 1081 HV Amsterdam, The Netherlands}}\\
\textsuperscript{63}{\footnotesize{University of Maryland, College Park, MD 20742, USA}}\\
\textsuperscript{64}{\footnotesize{Center for Relativistic Astrophysics and School of Physics, Georgia Institute of Technology, Atlanta, GA 30332, USA}}\\
\textsuperscript{65}{\footnotesize{Institut Lumi\`{e}re Mati\`{e}re, Universit\'{e} de Lyon, Universit\'{e} Claude Bernard Lyon 1, UMR CNRS 5306, F-69622 Villeurbanne, France}}\\
\textsuperscript{66}{\footnotesize{Laboratoire des Mat\'eriaux Avanc\'es (LMA), IN2P3/CNRS, Universit\'e de Lyon, F-69622 Villeurbanne, Lyon, France}}\\
\textsuperscript{67}{\footnotesize{Universitat de les Illes Balears, IAC3---IEEC, E-07122 Palma de Mallorca, Spain}}\\
\textsuperscript{68}{\footnotesize{Universit\`a di Napoli 'Federico II', Complesso Universitario di Monte S.Angelo, I-80126 Napoli, Italy}}\\
\textsuperscript{69}{\footnotesize{Canadian Institute for Theoretical Astrophysics, University of Toronto, Toronto, Ontario M5S 3H8, Canada}}\\
\textsuperscript{70}{\footnotesize{Tsinghua University, Beijing 100084, China}}\\
\textsuperscript{71}{\footnotesize{Texas Tech University, Lubbock, TX 79409, USA}}\\
\textsuperscript{72}{\footnotesize{The Pennsylvania State University, University Park, PA 16802, USA}}\\
\textsuperscript{73}{\footnotesize{National Tsing Hua University, Hsinchu City, 30013, Taiwan R.O.C.}}\\
\textsuperscript{74}{\footnotesize{Charles Sturt University, Wagga Wagga, New South Wales 2678, Australia}}\\
\textsuperscript{75}{\footnotesize{University of Chicago, Chicago, IL 60637, USA}}\\
\textsuperscript{76}{\footnotesize{Caltech CaRT, Pasadena, CA 91125, USA}}\\
\textsuperscript{77}{\footnotesize{Korea Institute of Science and Technology Information, Daejeon 305-806, Korea}}\\
\textsuperscript{78}{\footnotesize{Carleton College, Northfield, MN 55057, USA}}\\
\textsuperscript{79}{\footnotesize{Universit\`a di Roma ``La Sapienza,'' I-00185 Roma, Italy}}\\
\textsuperscript{80}{\footnotesize{University of Brussels, Brussels B-1050, Belgium}}\\
\textsuperscript{81}{\footnotesize{Sonoma State University, Rohnert Park, CA 94928, USA}}\\
\textsuperscript{82}{\footnotesize{Northwestern University, Evanston, IL 60208, USA}}\\
\textsuperscript{83}{\footnotesize{University of Minnesota, Minneapolis, MN 55455, USA}}\\
\textsuperscript{84}{\footnotesize{The University of Melbourne, Parkville, Victoria 3010, Australia}}\\
\textsuperscript{85}{\footnotesize{The University of Texas Rio Grande Valley, Brownsville, TX 78520, USA}}\\
\textsuperscript{86}{\footnotesize{The University of Sheffield, Sheffield S10 2TN, United Kingdom}}\\
\textsuperscript{87}{\footnotesize{University of Sannio at Benevento, I-82100 Benevento, Italy and INFN, Sezione di Napoli, I-80100 Napoli, Italy}}\\
\textsuperscript{88}{\footnotesize{Montclair State University, Montclair, NJ 07043, USA}}\\
\textsuperscript{89}{\footnotesize{Universit\`a di Trento, Dipartimento di Fisica, I-38123 Povo, Trento, Italy}}\\
\textsuperscript{90}{\footnotesize{INFN, Trento Institute for Fundamental Physics and Applications, I-38123 Povo, Trento, Italy}}\\
\textsuperscript{91}{\footnotesize{Cardiff University, Cardiff CF24 3AA, United Kingdom}}\\
\textsuperscript{92}{\footnotesize{National Astronomical Observatory of Japan, 2-21-1 Osawa, Mitaka, Tokyo 181-8588, Japan}}\\
\textsuperscript{93}{\footnotesize{School of Mathematics, University of Edinburgh, Edinburgh EH9 3FD, United Kingdom}}\\
\textsuperscript{94}{\footnotesize{Indian Institute of Technology, Gandhinagar, Ahmedabad, Gujarat 382424, India}}\\
\textsuperscript{95}{\footnotesize{Institute for Plasma Research, Bhat, Gandhinagar 382428, India}}\\
\textsuperscript{96}{\footnotesize{University of Szeged, D\'om t\'er 9, Szeged 6720, Hungary}}\\
\textsuperscript{97}{\footnotesize{Embry-Riddle Aeronautical University, Prescott, AZ 86301, USA}}\\
\textsuperscript{98}{\footnotesize{University of Michigan, Ann Arbor, MI 48109, USA}}\\
\textsuperscript{99}{\footnotesize{Tata Institute of Fundamental Research, Mumbai 400005, India}}\\
\textsuperscript{100}{\footnotesize{IISER-TVM, CET Campus, Trivandrum Kerala 695016, India}}\\
\textsuperscript{101}{\footnotesize{American University, Washington, D.C. 20016, USA}}\\
\textsuperscript{102}{\footnotesize{University of Massachusetts-Amherst, Amherst, MA 01003, USA}}\\
\textsuperscript{103}{\footnotesize{University of Adelaide, Adelaide, South Australia 5005, Australia}}\\
\textsuperscript{104}{\footnotesize{West Virginia University, Morgantown, WV 26506, USA}}\\
\textsuperscript{105}{\footnotesize{University of Bia{\l }ystok, 15-424 Bia{\l }ystok, Poland}}\\
\textsuperscript{106}{\footnotesize{SUPA, University of Strathclyde, Glasgow G1 1XQ, United Kingdom}}\\
\textsuperscript{107}{\footnotesize{Institute of Applied Physics, Nizhny Novgorod, 603950, Russia}}\\
\textsuperscript{108}{\footnotesize{Pusan National University, Busan 609-735, Korea}}\\
\textsuperscript{109}{\footnotesize{Hanyang University, Seoul 133-791, Korea}}\\
\textsuperscript{110}{\footnotesize{NCBJ, 05-400 \'Swierk-Otwock, Poland}}\\
\textsuperscript{111}{\footnotesize{IM-PAN, 00-956 Warsaw, Poland}}\\
\textsuperscript{112}{\footnotesize{Rochester Institute of Technology, Rochester, NY 14623, USA}}\\
\textsuperscript{113}{\footnotesize{Monash University, Victoria 3800, Australia}}\\
\textsuperscript{114}{\footnotesize{Seoul National University, Seoul 151-742, Korea}}\\
\textsuperscript{115}{\footnotesize{University of Alabama in Huntsville, Huntsville, AL 35899, USA}}\\
\textsuperscript{116}{\footnotesize{ESPCI, CNRS, F-75005 Paris, France}}\\
\textsuperscript{117}{\footnotesize{Universit\`a di Camerino, Dipartimento di Fisica, I-62032 Camerino, Italy}}\\
\textsuperscript{118}{\footnotesize{Southern University and A\&M College, Baton Rouge, LA 70813, USA}}\\
\textsuperscript{119}{\footnotesize{College of William and Mary, Williamsburg, VA 23187, USA}}\\
\textsuperscript{120}{\footnotesize{Instituto de F\'\i sica Te\'orica, University Estadual Paulista/ICTP South American Institute for Fundamental Research, S\~ao Paulo SP 01140-070, Brazil}}\\
\textsuperscript{121}{\footnotesize{University of Cambridge, Cambridge CB2 1TN, United Kingdom}}\\
\textsuperscript{122}{\footnotesize{IISER-Kolkata, Mohanpur, West Bengal 741252, India}}\\
\textsuperscript{123}{\footnotesize{Rutherford Appleton Laboratory, HSIC, Chilton, Didcot, Oxon OX11 0QX, United Kingdom}}\\
\textsuperscript{124}{\footnotesize{Whitman College, 345 Boyer Ave, Walla Walla, WA 99362  USA}}\\
\textsuperscript{125}{\footnotesize{National Institute for Mathematical Sciences, Daejeon 305-390, Korea}}\\
\textsuperscript{126}{\footnotesize{Hobart and William Smith Colleges, Geneva, NY 14456, USA}}\\
\textsuperscript{127}{\footnotesize{Janusz Gil Institute of Astronomy, University of Zielona G\'ora, 65-265 Zielona G\'ora,  Poland}}\\
\textsuperscript{128}{\footnotesize{Andrews University, Berrien Springs, MI 49104, USA}}\\
\textsuperscript{129}{\footnotesize{Universit\`a di Siena, I-53100 Siena, Italy}}\\
\textsuperscript{130}{\footnotesize{Trinity University, San Antonio, TX 78212, USA}}\\
\textsuperscript{131}{\footnotesize{University of Washington, Seattle, WA 98195, USA}}\\
\textsuperscript{132}{\footnotesize{Kenyon College, Gambier, OH 43022, USA}}\\
\textsuperscript{133}{\footnotesize{Abilene Christian University, Abilene, TX 79699, USA}}\\
\textsuperscript{134}{\footnotesize{{CSIRO Astronomy and Space Science, PO Box 76, Epping NSW 1710, Australia}}}\\
\textsuperscript{135}{\footnotesize{{ARC Centre of Excellence for All-sky Astrophysics (CAASTRO)}}}\\
\textsuperscript{136}{\footnotesize{Cornell Center for Astrophysics and Planetary Science, Ithaca, NY 14853, USA}}\\
\textsuperscript{137}{\footnotesize{Department of Physics and Electronics, Rhodes University, PO Box 94, Grahamstown, 6140, South Africa}}\\
\textsuperscript{138}{\footnotesize{CSIRO Astronomy and Space Science, 26 Dick Perry Avenue, Technology Park, Kensington WA 6151, Australia}}\\
\textsuperscript{139}{\footnotesize{{Sydney Institute for Astronomy, School of Physics, The University of Sydney, NSW 2006, Australia}}}\\
\textsuperscript{140}{\footnotesize{International Centre for Radio Astronomy Research (ICRAR), The University of Western Australia, M468, 35 Stirling Highway, Crawley, Perth, WA, 6009, Australia}}\\
\textsuperscript{141}{\footnotesize{Instituto de Astrof\'isica de Andaluc\'ia (IAA-CSIC), P.O. Box 03004, E-18080 Granada, Spain}}\\
\textsuperscript{142}{\footnotesize{Departamento de Ingenier\'ia de Sistemas y Autom\'atica, Escuela de Ingenier\'ias, Universidad de M\'alaga, Unidad Asociada al CSIC, Dr. Pedro Ortiz Ramos, E-29071 M\'alaga, Spain}}\\
\textsuperscript{143}{\footnotesize{Astronomical Institute, Academy of Sciences of the Czech Republic 251 65 Ond\v{r}ejov, Czech Republic}}\\
\textsuperscript{144}{\footnotesize{Institute of Physics of the Czech Academy of Sciences, Na Slovance 1999/2, 182 21 Praha 8, Czech Republic}}\\
\textsuperscript{145}{\footnotesize{Nikolaev National University, Nikolska str. 24, 54030 Nikolaev, Ukraine}}\\
\textsuperscript{146}{\footnotesize{Facultad de Ciencias, Universidad de M\'alaga, Bulevard Louis Pasteur, E-29010 M\'alaga, Spain}}\\
\textsuperscript{147}{\footnotesize{Ense\~nanza Virtual y Laboratorios Tecnol\'ogicos, Universidad de M\'alaga, Jim\'enez Fraud 10, E-29071 M\'alaga, Spain}}\\
\textsuperscript{148}{\footnotesize{ISDEFE for the SMOS FOS (ESA-ESAC), E-28692 Villanueva de la Ca\~nada (Madrid), Spain}}\\
\textsuperscript{149}{\footnotesize{Czech Technical University, Faculty of Electrical Engineering, Dep. of Radioelectronics, Technick\'a 2 166 27 Praha, Czech Republic}}\\
\textsuperscript{150}{\footnotesize{Astronomical Institute of the Academy of Sciences, Bo\v{c}n\'{\i} II 1401, CZ-14100 Praha 4, Czech Republic}}\\
\textsuperscript{151}{\footnotesize{Estaci\'on de Sondeos Atmosf\'ericos (ESAt) de El Arenosillo (CEDEA-INTA), E-21130 Mazag\'on, Huelva, Spain}}\\
\textsuperscript{152}{\footnotesize{Departamento de Ingenier\'ia Electr\'onica, Sistemas Inform\'aticos y Autom\'atica, Universidad de Huelva, E.T.S.I. de La R\'abida, E-21819 Palos de la Frontera (Huelva), Spain}}\\
\textsuperscript{153}{\footnotesize{Instituto de Hortofruticultura Subtropical y Mediterr\'anea La Mayora (IHSM/UMA-CSIC), E-29750 Algarrobo Costa (M\'alaga), Spain}}\\
\textsuperscript{154}{\footnotesize{Department of Physics, University of Auckland, Private Bag 92019, New Zealand}}\\
\textsuperscript{155}{\footnotesize{Vintage Lane Observatory, RD3, 7273 Blenheim, New Zealand}}\\
\textsuperscript{156}{\footnotesize{National Institute of Water and Atmospheric Research (NIWA), Lauder, New Zealand}}\\
\textsuperscript{157}{\footnotesize{Department of Physics, Sungkyunkwan University (SKKU), Suwon, Korea}}\\
\textsuperscript{158}{\footnotesize{Yunnan Astronomical Observatory, CAS, Kunming 650011, Yunnan, China}}\\
\textsuperscript{159}{\footnotesize{National Astronomical Observatory, Chinese Academy of Sciences, Beijing 100012, China}}\\
\textsuperscript{160}{\footnotesize{Instituto de Astronom\'ia, Universidad Nacional Aut\'onoma de M\'exico, 22800 Ensenada, Baja California, M\'exico}}\\
\textsuperscript{161}{\footnotesize{Instituto de Astronom\'ia, Universidad Nacional Aut\'onoma de M\'exico, Apdo Postal 70-264, Cd. Universitaria, 04510 M\'exico DF, M\'exico}}\\
\textsuperscript{162}{\footnotesize{Aryabhatta Research Institute of Observational Sciences, Manora Peak, Nainital-263 002, India}}\\
\textsuperscript{163}{\footnotesize{Escuela Polit\'ecnica Superior, Universidad de C\'adiz, Avda. Ram\'on Puyol, E-11202 Algeciras (C\'adiz), Spain}}\\
\textsuperscript{164}{\footnotesize{Divisi\'on de Ciencias del Espacio, Instituto Nacional de T\'ecnica Aerospacial (INTA), E-28850 Torrej\'on de Ardoz (Madrid), Spain}}\\
\textsuperscript{165}{\footnotesize{{Cerro Tololo Inter-American Observatory, National Optical Astronomy Observatory, Casilla 603, La Serena, Chile}}}\\
\textsuperscript{166}{\footnotesize{{Department of Physics \& Astronomy, University College London, Gower Street, London, WC1E 6BT, UK}}}\\
\textsuperscript{167}{\footnotesize{{Fermi National Accelerator Laboratory, P. O. Box 500, Batavia, IL 60510, USA}}}\\
\textsuperscript{168}{\footnotesize{{Department of Astrophysical Sciences, Princeton University, Peyton Hall, Princeton, NJ 08544, USA}}}\\
\textsuperscript{169}{\footnotesize{{CNRS, UMR 7095, Institut d'Astrophysique de Paris, F-75014, Paris, France}}}\\
\textsuperscript{170}{\footnotesize{{Sorbonne Universit\'es, UPMC Univ Paris 06, UMR 7095, Institut d'Astrophysique de Paris, F-75014, Paris, France}}}\\
\textsuperscript{171}{\footnotesize{Harvard-Smithsonian Center for Astrophysics, 60 Garden St, Cambridge, MA 02138, United States}}\\
\textsuperscript{172}{\footnotesize{{Carnegie Observatories, 813 Santa Barbara St., Pasadena, CA 91101, USA}}}\\
\textsuperscript{173}{\footnotesize{{Department of Physics and Astronomy, University of Pennsylvania, Philadelphia, PA 19104, USA}}}\\
\textsuperscript{174}{\footnotesize{{Kavli Institute for Particle Astrophysics \& Cosmology, P. O. Box 2450, Stanford University, Stanford, CA 94305, USA}}}\\
\textsuperscript{175}{\footnotesize{{SLAC National Accelerator Laboratory, Menlo Park, CA 94025, USA}}}\\
\textsuperscript{176}{\footnotesize{{Institute of Cosmology \& Gravitation, University of Portsmouth, Portsmouth, PO1 3FX, UK}}}\\
\textsuperscript{177}{\footnotesize{{Institut de Ci\`encies de l'Espai, IEEC-CSIC, Campus UAB, Carrer de Can Magrans, s/n, E-08193 Bellaterra, Barcelona, Spain}}}\\
\textsuperscript{178}{\footnotesize{{Institut de F\'{\i}sica d'Altes Energies (IFAE), The Barcelona Institute of Science and Technology, Campus UAB, E-08193 Bellaterra (Barcelona) Spain}}}\\
\textsuperscript{179}{\footnotesize{{Astrophysical Institute, Department of Physics and Astronomy, 251B Clippinger Lab, Ohio University, Athens, OH 45701, USA}}}\\
\textsuperscript{180}{\footnotesize{{Laborat\'orio Interinstitucional de e-Astronomia---LIneA, Rua Gal. Jos\'e Cristino 77, Rio de Janeiro, RJ - 20921-400, Brazil}}}\\
\textsuperscript{182}{\footnotesize{{Excellence Cluster Universe, Boltzmannstr.\ 2, D-85748 Garching, Germany}}}\\
\textsuperscript{183}{\footnotesize{{Ludwig Maximillian Universit\"at Munich, Faculty of Physics, Schellingstrasse 4, D-80799 Munich, Germany}}}\\
\textsuperscript{184}{\footnotesize{{Kavli Institute for Cosmological Physics, University of Chicago, Chicago, IL 60637, USA}}}\\
\textsuperscript{185}{\footnotesize{Jet Propulsion Laboratory, California Institute of Technology, Pasadena, CA 91109, USA}}\\
\textsuperscript{186}{\footnotesize{{Department of Astronomy, University of Illinois, 1002 W. Green Street, Urbana, IL 61801, USA}}}\\
\textsuperscript{187}{\footnotesize{{Department of Physics, University of Illinois, 1110 W. Green St., Urbana, IL 61801, USA}}}\\
\textsuperscript{188}{\footnotesize{{University of Arizona, Steward Observatory, University of Arizona, 933 N. Cherry Avenue, Tucson, AZ 85721, USA}}}\\
\textsuperscript{189}{\footnotesize{{Department of Astronomy \& Astrophysics, Center for Particle \& Gravitational Astrophysics, and Center for Theoretical \& Observational Cosmology, Pennsylvania State University, University Park, PA 16802, USA}}}\\
\textsuperscript{190}{\footnotesize{{CCS Division, Los Alamos National Laboratory, Los Alamos, NM 87545, USA}}}\\
\textsuperscript{191}{\footnotesize{{Department of Physics, University of Michigan, Ann Arbor, MI 48109, USA}}}\\
\textsuperscript{192}{\footnotesize{{Department of Astronomy, University of California, Berkeley,  501 Campbell Hall, Berkeley, CA 94720, USA}}}\\
\textsuperscript{193}{\footnotesize{{Lawrence Berkeley National Laboratory, 1 Cyclotron Road, Berkeley, CA 94720, USA}}}\\
\textsuperscript{194}{\footnotesize{{National Center for Supercomputing Applications, 1205 West Clark St., Urbana, IL 61801, USA}}}\\
\textsuperscript{195}{\footnotesize{{Center for Cosmology and Astro-Particle Physics, The Ohio State University, Columbus, OH 43210, USA}}}\\
\textsuperscript{196}{\footnotesize{{Department of Physics, The Ohio State University, Columbus, OH 43210, USA}}}\\
\textsuperscript{197}{\footnotesize{{Departments of Physics and Astronomy, University of California, Berkeley, CA, USA}}}\\
\textsuperscript{198}{\footnotesize{{Australian Astronomical Observatory, North Ryde, NSW 2113, Australia}}}\\
\textsuperscript{199}{\footnotesize{{George P. and Cynthia Woods Mitchell Institute for Fundamental Physics and Astronomy, and Department of Physics and Astronomy, Texas A\&M University, College Station, TX 77843,  USA}}}\\
\textsuperscript{200}{\footnotesize{{Departamento de F\'{\i}sica Matem\'atica,  Instituto de F\'{\i}sica, Universidade de S\~ao Paulo,  CP 66318, CEP 05314-970, S\~ao Paulo, SP,  Brazil}}}\\
\textsuperscript{201}{\footnotesize{{Center for Cosmology and Particle Physics, New York University, 4 Washington Place, New York, NY 10003, USA}}}\\
\textsuperscript{202}{\footnotesize{{Department of Astronomy, The Ohio State University, Columbus, OH 43210, USA}}}\\
\textsuperscript{203}{\footnotesize{{National Optical Astronomy Observatory, 950 North Cherry Avenue, Tucson, AZ 85719, USA}}}\\
\textsuperscript{204}{\footnotesize{{Columbia Astrophysics Laboratory, Pupin Hall, New York, NY, 10027,USA}}}\\
\textsuperscript{205}{\footnotesize{{Department of Astronomy, University of Michigan, Ann Arbor, MI 48109, USA}}}\\
\textsuperscript{206}{\footnotesize{{Instituci\'o Catalana de Recerca i Estudis Avan\c{c}ats, E-08010 Barcelona, Spain}}}\\
\textsuperscript{207}{\footnotesize{{Department of Astronomy \& Theoretical Astrophysics Center, University of California, Berkeley, CA 94720-3411, USA}}}\\
\textsuperscript{208}{\footnotesize{{Department of Physics and Astronomy, Pevensey Building, University of Sussex, Brighton, BN1 9QH, UK}}}\\
\textsuperscript{209}{\footnotesize{{Centro de Investigaciones Energ\'eticas, Medioambientales y Tecnol\'ogicas (CIEMAT), Madrid, Spain}}}\\
\textsuperscript{210}{\footnotesize{{Brookhaven National Laboratory, Bldg 510, Upton, NY 11973, USA}}}\\
\textsuperscript{211}{\footnotesize{{Argonne National Laboratory, 9700 South Cass Avenue, Lemont, IL 60439, USA}}}\\
\textsuperscript{212}{\footnotesize{{Jodrell Bank Center for Astrophysics, School of Physics and Astronomy, University of Manchester, Oxford Road, Manchester, M13 9PL, UK}}}\\
\textsuperscript{213}{\footnotesize{{Universities Space Research Association, 320 Sparkman Dr. Huntsville, AL 35806, USA}}}\\
\textsuperscript{214}{\footnotesize{{Physics Dept, University of Alabama in Huntsville, 320 Sparkman Dr., Huntsville, AL 35899, USA}}}\\
\textsuperscript{215}{\footnotesize{{NASA Postdoctoral Program Fellow}}}\\
\textsuperscript{216}{\footnotesize{{Astrophysics Office, ZP12, NASA/Marshall Space Flight Center, Huntsville, AL 35812, USA}}}\\
\textsuperscript{217}{\footnotesize{{Dept. of Space Science, University of Alabama in Huntsville, 320 Sparkman Dr., Huntsville, AL 35899, USA}}}\\
\textsuperscript{218}{\footnotesize{{CSPAR, University of Alabama in Huntsville, 320 Sparkman Dr., Huntsville, AL 35899, USA}}}\\
\textsuperscript{219}{\footnotesize{{Instituto de Astrof\'isica de Andaluc\'a (IAA-CSIC), P.O. Box 03004, E-18080 Granada, Spain}}}\\
\textsuperscript{220}{\footnotesize{{Istituto Nazionale di Fisica Nucleare, Sezione di Bari, I-70126 Bari, Italy}}}\\
\textsuperscript{221}{\footnotesize{{Jacobs Technology, Inc., Huntsville, AL, USA}}}\\
\textsuperscript{222}{\footnotesize{Max-Planck-Institut f{\"u}r Extraterrestrische Physik, Giessenbachstra\ss e 1, D-85748, Garching, Germany}}\\
\textsuperscript{223}{\footnotesize{{Los Alamos National Laboratory, NM 87545, USA}}}\\
\textsuperscript{224}{\footnotesize{{School of Physics, University College Dublin, Belfield, Stillorgan Road, Dublin 4, Ireland}}}\\
\textsuperscript{225}{\footnotesize{{NASA Headquarters, Washington DC, USA}}}\\
\textsuperscript{226}{\footnotesize{{Excellence Cluster Universe, Technische Universit\"at M\"unchen, Boltzmannstr. 2, D-85748, Garching, Germany}}}\\
\textsuperscript{227}{\footnotesize{{Deutsches Elektronen Synchrotron DESY, D-15738 Zeuthen, Germany}}}\\
\textsuperscript{228}{\footnotesize{{Department of Physics and Astronomy, Clemson University, Kinard Lab of Physics, Clemson, SC 29634-0978, USA}}}\\
\textsuperscript{229}{\footnotesize{{W. W. Hansen Experimental Physics Laboratory, Kavli Institute for Particle Astrophysics and Cosmology, Department of Physics and SLAC National Accelerator Laboratory, Stanford University, Stanford, CA 94305, USA}}}\\
\textsuperscript{230}{\footnotesize{{Department of Physics, Stockholm University, AlbaNova, SE-106 91 Stockholm, Sweden}}}\\
\textsuperscript{231}{\footnotesize{{The Oskar Klein Centre for Cosmoparticle Physics, AlbaNova, SE-106 91 Stockholm, Sweden}}}\\
\textsuperscript{232}{\footnotesize{{Santa Cruz Institute for Particle Physics, Department of Physics and Department of Astronomy and Astrophysics, University of California at Santa Cruz, Santa Cruz, CA 95064, USA}}}\\
\textsuperscript{233}{\footnotesize{{Department of Physics, KTH Royal Institute of Technology, AlbaNova, SE-106 91 Stockholm, Sweden}}}\\
\textsuperscript{234}{\footnotesize{{Tokyo Metropolitan University, Department of Physics, Minami-osawa 1-1, Hachioji, Tokyo 192-0397, Japan}}}\\
\textsuperscript{235}{\footnotesize{{Universit\`a di Pisa and Istituto Nazionale di Fisica Nucleare, Sezione di Pisa I-56127 Pisa, Italy}}}\\
\textsuperscript{236}{\footnotesize{{Istituto Nazionale di Fisica Nucleare, Sezione di Trieste, I-34127 Trieste, Italy}}}\\
\textsuperscript{237}{\footnotesize{{Dipartimento di Fisica, Universit\`a di Trieste, I-34127 Trieste, Italy}}}\\
\textsuperscript{238}{\footnotesize{{Istituto Nazionale di Fisica Nucleare, Sezione di Padova, I-35131 Padova, Italy}}}\\
\textsuperscript{239}{\footnotesize{{Dipartimento di Fisica e Astronomia ``G. Galilei'', Universit\`a di Padova, I-35131 Padova, Italy}}}\\
\textsuperscript{240}{\footnotesize{{Istituto Nazionale di Fisica Nucleare, Sezione di Pisa, I-56127 Pisa, Italy}}}\\
\textsuperscript{241}{\footnotesize{{Istituto Nazionale di Fisica Nucleare, Sezione di Bari, I-70126 Bari, Italy}}}\\
\textsuperscript{242}{\footnotesize{{Istituto Nazionale di Fisica Nucleare, Sezione di Torino, I-10125 Torino, Italy}}}\\
\textsuperscript{243}{\footnotesize{{Dipartimento di Fisica Generale ``Amadeo Avogadro" , Universit\`a degli Studi di Torino, I-10125 Torino, Italy}}}\\
\textsuperscript{244}{\footnotesize{{NASA Goddard Space Flight Center, Greenbelt, MD 20771, USA}}}\\
\textsuperscript{245}{\footnotesize{{Laboratoire Leprince-Ringuet, \'Ecole polytechnique, CNRS/IN2P3, Palaiseau, France}}}\\
\textsuperscript{246}{\footnotesize{{Department of Physics and Center for Space Sciences and Technology, University of Maryland Baltimore County, Baltimore, MD 21250, USA}}}\\
\textsuperscript{247}{\footnotesize{{Center for Research and Exploration in Space Science and Technology (CRESST) and NASA Goddard Space Flight Center, Greenbelt, MD 20771, USA}}}\\
\textsuperscript{248}{\footnotesize{{Consorzio Interuniversitario per la Fisica Spaziale (CIFS), I-10133 Torino, Italy}}}\\
\textsuperscript{249}{\footnotesize{{Dipartimento di Fisica ``M. Merlin" dell'Universit\`a e del Politecnico di Bari, I-70126 Bari, Italy}}}\\
\textsuperscript{250}{\footnotesize{{INAF---Istituto di Astrofisica Spaziale e Fisica Cosmica, I-20133 Milano, Italy}}}\\
\textsuperscript{251}{\footnotesize{{Agenzia Spaziale Italiana (ASI) Science Data Center, I-00133 Roma, Italy}}}\\
\textsuperscript{252}{\footnotesize{{College of Science, George Mason University, Fairfax, VA 22030, resident at Naval Research Laboratory, Washington, DC 20375, USA}}}\\
\textsuperscript{253}{\footnotesize{{Istituto Nazionale di Fisica Nucleare, Sezione di Perugia, I-06123 Perugia, Italy}}}\\
\textsuperscript{254}{\footnotesize{{Laboratoire Univers et Particules de Montpellier, Universit\'e Montpellier, CNRS/IN2P3, Montpellier, France}}}\\
\textsuperscript{255}{\footnotesize{{Department of Physics and Astronomy, Sonoma State University, Rohnert Park, CA 94928-3609, USA}}}\\
\textsuperscript{256}{\footnotesize{{INAF Istituto di Radioastronomia, I-40129 Bologna, Italy}}}\\
\textsuperscript{257}{\footnotesize{{Dipartimento di Astronomia, Universit\`a di Bologna, I-40127 Bologna, Italy}}}\\
\textsuperscript{258}{\footnotesize{{Universit\`a Telematica Pegaso, Piazza Trieste e Trento, 48, I-80132 Napoli, Italy}}}\\
\textsuperscript{259}{\footnotesize{{Universit\`a di Udine, I-33100 Udine, Italy}}}\\
\textsuperscript{260}{\footnotesize{{Department of Physical Sciences, Hiroshima University, Higashi-Hiroshima, Hiroshima 739-8526, Japan}}}\\
\textsuperscript{261}{\footnotesize{{Erlangen Centre for Astroparticle Physics, D-91058 Erlangen, Germany}}}\\
\textsuperscript{262}{\footnotesize{{Instituto de Astrof\'isica, Facultad de F\'isica, Pontificia Universidad Cat\'olica de Chile, Casilla 306, Santiago 22, Chile}}}\\
\textsuperscript{263}{\footnotesize{{Department of Physics and Department of Astronomy, University of Maryland, College Park, MD 20742, USA}}}\\
\textsuperscript{264}{\footnotesize{{Laboratoire AIM, CEA-IRFU/CNRS/Universit\'e Paris Diderot, Service d'Astrophysique, CEA Saclay, F-91191 Gif sur Yvette, France}}}\\
\textsuperscript{265}{\footnotesize{{Space Science Division, Naval Research Laboratory, Washington, DC 20375-5352, USA}}}\\
\textsuperscript{266}{\footnotesize{{NASA Postdoctoral Program Fellow, USA}}}\\
\textsuperscript{267}{\footnotesize{{Institut f\"ur Astro- und Teilchenphysik and Institut f\"ur Theoretische Physik, Leopold-Franzens-Universit\"at Innsbruck, A-6020 Innsbruck, Austria}}}\\
\textsuperscript{268}{\footnotesize{{University of North Florida, Department of Physics, 1 UNF Drive, Jacksonville, FL 32224 , USA}}}\\
\textsuperscript{269}{\footnotesize{{School of Physics and Astronomy, University of Southampton, Highfield, Southampton, SO17 1BJ, UK}}}\\
\textsuperscript{270}{\footnotesize{{Science Institute, University of Iceland, IS-107 Reykjavik, Iceland}}}\\
\textsuperscript{271}{\footnotesize{{Institute of Space Sciences (IEEC-CSIC), Campus UAB, E-08193 Barcelona, Spain}}}\\
\textsuperscript{272}{\footnotesize{{email: Julie.E.McEnery@nasa.gov}}}\\
\textsuperscript{273}{\footnotesize{{Hiroshima Astrophysical Science Center, Hiroshima University, Higashi-Hiroshima, Hiroshima 739-8526, Japan}}}\\
\textsuperscript{274}{\footnotesize{{Max-Planck-Institut f\"ur Physik, D-80805 M\"unchen, Germany}}}\\
\textsuperscript{275}{\footnotesize{{Department of Physics and Astronomy, University of Denver, Denver, CO 80208, USA}}}\\
\textsuperscript{276}{\footnotesize{{Department of Physics, University of Johannesburg, PO Box 524, Auckland Park 2006, South Africa}}}\\
\textsuperscript{277}{\footnotesize{{Department of Physics, The University of Hong Kong, Pokfulam Road, Hong Kong, China}}}\\
\textsuperscript{278}{\footnotesize{{The University of Hong Kong, Laboratory for Space Research, Hong Kong, China}}}\\
\textsuperscript{279}{\footnotesize{{NYCB Real-Time Computing Inc., Lattingtown, NY 11560-1025, USA}}}\\
\textsuperscript{280}{\footnotesize{{Department of Chemistry and Physics, Purdue University Calumet, Hammond, IN 46323-2094, USA}}}\\
\textsuperscript{281}{\footnotesize{{Solar-Terrestrial Environment Laboratory, Nagoya University, Nagoya 464-8601, Japan}}}\\
\textsuperscript{282}{\footnotesize{{Max-Planck-Institut f\"ur Kernphysik, D-69029 Heidelberg, Germany}}}\\
\textsuperscript{283}{\footnotesize{{Instituci\'o Catalana de Recerca i Estudis Avan\c{c}ats (ICREA), Barcelona, Spain}}}\\
\textsuperscript{284}{\footnotesize{{Department of Physics, 3-34-1 Nishi-Ikebukuro, Toshima-ku, Tokyo 171-8501, Japan}}}\\
\textsuperscript{285}{\footnotesize{INAF---Osservatorio Astronomico di Roma, via Frascati 33, I-00078 Monte Porzio Catone (RM), Italy}}\\
\textsuperscript{286}{\footnotesize{INAF---Osservatorio Astronomico di Padova, Vicolo Osservatorio 5,  I-35122 Padova, Italy}}\\
\textsuperscript{287}{\footnotesize{INAF---Osservatorio Astronomico di Brera, via E. Bianchi 46, I-23807 Merate, Italy}}\\
\textsuperscript{288}{\footnotesize{INAF---Osservatorio Astronomico di Capodimonte, salita Moiariello 16, I-80131 Napoli, Italy}}\\
\textsuperscript{289}{\footnotesize{{INAF---Istituto di Astrofisica Spaziale e Fisica Cosmica di Bologna, via Gobetti 101, I-40129 Bologna, Italy}}}\\
\textsuperscript{290}{\footnotesize{{Scuola Normale Superiore, Piazza dei Cavalieri, 7, I-56126 Pisa, Italy}}}\\
\textsuperscript{291}{\footnotesize{ASI-Science Data Center, via dl Politecnico s.n.c., I-00133 Roma, Italy}}\\
\textsuperscript{292}{\footnotesize{Dip. di Fisica Ettore Pancini, University of Naples “Federico II”, C.U. Monte Sant'Angelo, Via Cinthia, I-80126, Napoli, Italy}}\\
\textsuperscript{293}{\footnotesize{{INAF---ORA---Osservatorio Astronomico di Cagliari, Via della Scienza n. 5, I-09047 Selargius (CA), Italy}}}\\
\textsuperscript{294}{\footnotesize{{INAF---Osservatorio Astronomico di Torino, Strada Osservatorio 20, I-10025, Pino Torinese (To), Italy}}}\\
\textsuperscript{295}{\footnotesize{{INAF---Institute for Space Astrophysics and Planetology, Via Fosso del Cavaliere 100, I-00133 Rome, Italy}}}\\
\textsuperscript{296}{\footnotesize{{ISDC, Department of astronomy, University of Geneva, chemin d'\'Ecogia, 16 CH-1290 Versoix, Switzerland}}}\\
\textsuperscript{297}{\footnotesize{{DTU Space---National Space Institute Elektrovej---Building 327 DK-2800 Kongens Lyngby Denmark}}}\\
\textsuperscript{298}{\footnotesize{{Space Science Group, School of Physics, University College Dublin, Belfield, Dublin 4, Ireland}}}\\
\textsuperscript{299}{\footnotesize{{European Space Astronomy Centre (ESA/ESAC), Science Operations Department, E-28691, Villanueva de la Ca\ ̃nada, Madrid, Spain}}}\\
\textsuperscript{300}{\footnotesize{{APC, AstroParticule et Cosmologie, Universit\'e Paris Diderot, CNRS/IN2P3, CEA/Irfu, Observatoire de Paris, Sorbonne Paris Cit\'e, 10 rue Alice Domont et L\'eonie Duquet, F-75205 Paris Cedex 13, France}}}\\
\textsuperscript{301}{\footnotesize{{INAF,  IASF-Milano, via E.Bassini 15, I-20133 Milano, Italy}}}\\
\textsuperscript{302}{\footnotesize{{Universit\'e Toulouse; UPS-OMP; CNRS; IRAP; 9 Av. Roche, BP 44346, F-31028 Toulouse, France}}}\\
\textsuperscript{303}{\footnotesize{{Fran\c{c}ois Arago Centre, APC, Universit\'e Paris Diderot, CNRS/IN2P3, CEA/Irfu, Observatoire de Paris,, Sorbonne Paris Cit\'e, 10 rue Alice Domon et L\'eonie Duquet, 75205 Paris Cedex 13, France}}}\\
\textsuperscript{304}{\footnotesize{Cahill Center for Astrophysics, California Institute of Technology, Pasadena, CA 91125, USA}}\\
\textsuperscript{305}{\footnotesize{{Hubble Fellow}}}\\
\textsuperscript{306}{\footnotesize{Department of Particle Physics and Astrophysics, Weizmann Institute of Science, 76100 Rehovot, Israel}}\\
\textsuperscript{307}{\footnotesize{Infrared Processing and Analysis Center, California Institute of Technology, Pasadena, CA 91125, USA}}\\
\textsuperscript{308}{\footnotesize{Max Planck Institute for Astronomy, Königstuhl 17, D-69117 Heidelberg, Germany}}\\
\textsuperscript{309}{\footnotesize{Dark Cosmology Centre, Niels Bohr Institute, Juliane Maries Vej 30, Copenhagen Ø, DK-2100, Denmark}}\\
\textsuperscript{310}{\footnotesize{Department of Astronomy and the Oskar Klein Centre, Stockholm University, AlbaNova, SE-106 91 Stockholm, Sweden}}\\
\textsuperscript{311}{\footnotesize{Lockheed Martin Space Systems Company, Palo Alto, CA 94304}}\\
\textsuperscript{312}{\footnotesize{{University of California, Berkeley, Space Sciences Laboratory, 7 Gauss Way, Berkeley, CA 94720-7450, USA}}}\\
\textsuperscript{313}{\footnotesize{{Ioffe Physical Technical Institute, Politekhnicheskaya 26, St. Petersburg, 194021, Russia}}}\\
\textsuperscript{314}{\footnotesize{{Emeritus}}}\\
\textsuperscript{315}{\footnotesize{{Universities Space Research Association, 7178 Columbia Gateway Drive, Columbia, MD 21046 USA }}}\\
\textsuperscript{316}{\footnotesize{{Institute for Space-Earth Environmental Research, Nagoya University, Furo-cho, Chikusa-ku, Nagoya 464-8601, Japan}}}\\
\textsuperscript{317}{\footnotesize{{Institute of Astronomy, Graduate School of Science, The University of Tokyo, Mitaka, Tokyo 181-0015, Japan}}}\\
\textsuperscript{318}{\footnotesize{{The Research Institute for Time Studies, Yamaguchi University, Yamaguchi, Yamaguchi 753-8511, Japan}}}\\
\textsuperscript{319}{\footnotesize{{Hiroshima Astrophysical Science Center, Hiroshima University, Higashi-Hiroshima, Hiroshima 739-8526, Japan}}}\\
\textsuperscript{320}{\footnotesize{{Division of Theoretical Astronomy, National Astronomical Observatory of Japan, Mitaka, Tokyo 181-8588, Japan}}}\\
\textsuperscript{321}{\footnotesize{{Department of Astronomy, Kyoto University, Kyoto, Kyoto 606-8502, Japan}}}\\
\textsuperscript{322}{\footnotesize{{Okayama Astrophysical Observatory, National Astronomical Observatory of Japan, Asakuchi, Okayama 719-0232, Japan}}}\\
\textsuperscript{323}{\footnotesize{Physics Department, Yale University, New Haven, CT 06520, USA}}\\
\textsuperscript{324}{\footnotesize{Astrophysics Research Institue, Liverpool JMU, Liverpool L3 5RF, UK}}\\
\textsuperscript{325}{\footnotesize{{Monash Centre for Astrophysics (MoCA), Monash University, Clayton VIC 3800, Australia}}}\\
\textsuperscript{326}{\footnotesize{{School of Physics \& Astronomy, Monash University, Clayton VIC 3800, Australia}}}\\
\textsuperscript{327}{\footnotesize{University of Nova Gorica, Vipavska 13, 5000 Nova Gorica, Slovenia}}\\
\textsuperscript{328}{\footnotesize{Faculty of Mathematics and Physics, University of Ljubljana, Jadranska 19, 1000 Ljubljana, Slovenia}}\\
\textsuperscript{329}{\footnotesize{Department of Physics, University of Bath, BA2 7AY, UK}}\\
\textsuperscript{330}{\footnotesize{University of Warwick, Department of Physics, Gibbet Hill Road, Coventry, CV4 7AL, UK}}\\
\textsuperscript{331}{\footnotesize{{ASTRON, The Netherlands Institute for Radio Astronomy, Postbus 2, 7990 AA, Dwingeloo, The Netherlands}}}\\
\textsuperscript{332}{\footnotesize{{Astrophysics, Department of Physics, University of Oxford, Keble Road, Oxford OX1 3RH, UK}}}\\
\textsuperscript{333}{\footnotesize{{SRON Netherlands Institute for Space Research, Sorbonnelaan 2, 3584 CA Utrecht, the Netherlands}}}\\
\textsuperscript{334}{\footnotesize{{Anton Pannekoek Institute for Astronomy,  University of Amsterdam, Science Park 904, 1098 XH Amsterdam, The Netherlands}}}\\
\textsuperscript{335}{\footnotesize{Lomonosov Moscow State University, Sternberg Astronomical Institute, 13, Universitetskiy prospekt, Moscow, 119234, Russia}}\\
\textsuperscript{336}{\footnotesize{South African Astronomical Observatory, PO Box 9, 7935 Observatory , Cape Town, South Africa}}\\
\textsuperscript{337}{\footnotesize{The Instituto de Astrofisica de Canarias, Calle Via Lactea, s/n, E-38200 La Laguna, Tenerife, Spain}}\\
\textsuperscript{338}{\footnotesize{Applied Physics Institute, Irkutsk State University, 20, Gagarin blvd,, Irkutsk, 664003, Russia}}\\
\textsuperscript{339}{\footnotesize{Kislovodsk Solar Station of the Main (Pulkovo) Observatory RAS, P.O.Box 45, ul. Gagarina 100, Kislovodsk, 357700, Russia}}\\
\textsuperscript{340}{\footnotesize{Blagoveschensk State Pedagogical University, Lenin str., 104, Amur Region, Blagoveschensk, 675000, Russia}}\\
\textsuperscript{341}{\footnotesize{{Department of Physics, Tokyo Institute of Technology, Meguro-ku, Tokyo 152-8851, Japan}}}\\
\textsuperscript{342}{\footnotesize{{MAXI team, RIKEN, 2-1 Hirosawa, Wako, Saitama 351-0198, Japan}}}\\
\textsuperscript{343}{\footnotesize{{Department of Physics, Nihon University, 1-8-14 Kanda-Surugadai, Chiyoda-ku, Tokyo 101-8308, Japan}}}\\
\textsuperscript{344}{\footnotesize{{JEM Mission Operations and Integration Center, Human Spaceflight Technology Directorate,  Japan Aerospace Exploration Agency, 2-1-1 Sengen, Tsukuba, Ibaraki 305-8505, Japan}}}\\
\textsuperscript{345}{\footnotesize{{Institute of Space and Astronautical Science (ISAS), Japan Aerospace Exploration Agency (JAXA), 3-1-1 Yoshinodai, Chuo, Sagamihara, Kanagawa 252-5210, Japan}}}\\
\textsuperscript{346}{\footnotesize{{Department of Earth and Space Science, Osaka University, 1-1 Machikaneyama, Toyonaka, Osaka 560-0043, Japan}}}\\
\textsuperscript{347}{\footnotesize{{University of California, Berkeley, Astronomy Dept., 501 Campbell Hall \#3411, Berkeley, CA 94720, USA}}}\\
\textsuperscript{348}{\footnotesize{{Eureka Scientific, Inc., 2452 Delmer Street Suite 100, Oakland, CA 94602, USA}}}\\
\textsuperscript{349}{\footnotesize{{Kavli Institute for Astrophysics and Space Research, Massachusetts Institute of Technology, Cambridge, MA 02139, USA}}}\\
\textsuperscript{350}{\footnotesize{{International Centre for Radio Astronomy Research, Curtin University, Bentley, WA 6102, Australia}}}\\
\textsuperscript{351}{\footnotesize{{Dunlap Institute for Astronomy and Astrophysics, University of Toronto, Toronto, ON M5S 3H4, Canada}}}\\
\textsuperscript{352}{\footnotesize{{School of Chemical \& Physical Sciences, Victoria University of Wellington, PO Box 600,  Wellington 6140, New Zealand}}}\\
\textsuperscript{353}{\footnotesize{{Osservatorio di Radio Astronomia, Istituto Nazionale di Astrofisica, Bologna, I-40123, Italy}}}\\
\textsuperscript{354}{\footnotesize{Astrophysics Research Centre, School of Mathematics and Physics, Queens University Belfast, Belfast BT7 1NN, UK}}\\
\textsuperscript{355}{\footnotesize{Institute for Astronomy, University of Hawaii at Manoa, Honolulu, HI 96822, USA}}\\
\textsuperscript{356}{\footnotesize{Space Telescope Science Institute, 3700 San Martin Drive, Baltimore, MD 21218, USA}}\\
\textsuperscript{357}{\footnotesize{Department of Physics, Harvard University, Cambridge, MA 02138, USA}}\\
\textsuperscript{358}{\footnotesize{Millennium Institute of Astrophysics, Casilla 36–D, Santiago, Chile}}\\
\textsuperscript{359}{\footnotesize{Departamento de Ciencias Fisicas, Universidad Andres Bello, Avda. Republica 252, Santiago, Chile}}\\
\textsuperscript{360}{\footnotesize{Institute of Astronomy, University of Cambridge, Madingley Road, Cambridge CB3 0HA, UK}}\\
\textsuperscript{361}{\footnotesize{Institut d'Astrophysique de Paris, CNRS, and Universit\'e Pierre et Marie Curie, 98 bis Boulevard Arago, F-75014, Paris, France}}\\
\textsuperscript{362}{\footnotesize{European Southern Observatory, Alonso de Cordova 3107, Vitacura, Santiago, Chile}}\\
\textsuperscript{363}{\footnotesize{Tuorla Observatory, Department of Physics and Astronomy, University of Turku, V{\" a}i{\"a}l{\" a}ntie 20, FI-21500 Piikki{\" o}, Finland}}\\
\textsuperscript{364}{\footnotesize{Departamento de Astronomia, Universidad de Chile, Camino El Observatorio 1515, Las Condes, Santiago, Chile}}\\
\textsuperscript{365}{\footnotesize{Sorbonne Universit\'es, UPMC Univ. Paris 06, UMR 7585, LPNHE, F-75005, Paris, France}}\\
\textsuperscript{366}{\footnotesize{CNRS, UMR 7585, Laboratoire de Physique Nucleaire et des Hautes Energies, 4 place Jussieu, F-75005 Paris, France}}\\
\textsuperscript{367}{\footnotesize{Las Cumbres Observatory Global Telescope Network, 6740 Cortona Dr., Suite 102, Goleta, CA 93117, USA}}\\
\textsuperscript{368}{\footnotesize{Department of Physics, University of California Santa Barbara, Santa Barbara, CA 93106, USA}}\\
\textsuperscript{369}{\footnotesize{{Benoziyo Center for Astrophysics, Weizmann Institute of Science, 76100 Rehovot, Israel}}}\\
\textsuperscript{370}{\footnotesize{Faculty of Physics, University of Warsaw, 02-093 Warszawa, Poland}}\\
\textsuperscript{371}{\footnotesize{Center for Theoretical Physics of the Polish Academy of Sciences, 02-668 Warszawa, Poland}}\\
\textsuperscript{372}{\footnotesize{Department of Physics and Astronomy, University of Leicester, Leicester, LE1 7RH, UK}}\\
\textsuperscript{373}{\footnotesize{Joint Space-Science Institute, University of Maryland, College Park, MD 20742, USA}}\\
\textsuperscript{374}{\footnotesize{Los Alamos National Laboratory, B244, Los Alamos, NM, 87545, USA}}\\
\textsuperscript{375}{\footnotesize{L'Institut de Recherche en Astrophysique et Plan\'etologie, CNRS UMR 5277/UPS, 14 avenue Edouard Belin, F-31400 Toulouse, France}}\\
\textsuperscript{376}{\footnotesize{{University of Texas at San Antonio, San Antonio, TX, USA}}}\\
\textsuperscript{377}{\footnotesize{{Mitchell Institute for Fundamental Physics and Astronomy, Department of Physics and Astronomy, Texas A\&M University, 4242 TAMU, College Station, TX 77843, USA}}}\\
\textsuperscript{378}{\footnotesize{{Universidad Nacional de C\'ordoba, IATE, Laprida 854, C\'ordoba, Argentina}}}\\
\textsuperscript{379}{\footnotesize{{Instituto de Astrof\'isica, Pontificia Universidad Cat\'olica de Chile, Av. Vicu\~na Mackenna 4860, Santiago, Chile}}}\\
\textsuperscript{380}{\footnotesize{{Instituto de Astrof\'isica de Andaluc\'ia, Consejo Superior de Investigaciones Cient\'ificas, Glorieta de la Astronom\'ia s/n, E-18008 Granada, Spain}}}\\
\textsuperscript{381}{\footnotesize{{Centre for Astrophysics and Cosmology, Science Institute, University of Iceland, 107 Reykjavik, Iceland}}}\\
 
\end{document}